\documentclass[12pt,a4paper]{article}
\nonstopmode
\usepackage{epsfig}
\usepackage[a4paper]{geometry}
\usepackage{float}
\usepackage[stable]{footmisc}
\usepackage[utf8]{inputenc}
\usepackage{a4wide}
\usepackage{amsmath,amssymb,amstext,amsthm,amssymb}
\usepackage{amsfonts}
\usepackage{framed}
\usepackage{graphicx}
\usepackage{color}
\usepackage[dvipsnames]{xcolor}
\usepackage{bbold}
\usepackage{bbm}
\usepackage[normalem]{ulem}
\usepackage{rotating} 
\usepackage{slashed} 
\usepackage{cancel} 
\usepackage{braket} 
\usepackage{amstext}
\usepackage{array}
\usepackage{setspace}
\usepackage{hyperref}
\usepackage{overpic}
\usepackage{bbm}
\usepackage{cite}
\usepackage{ulem}
\usepackage{lipsum}
\usepackage{tikz}

\newcommand{\DeclareRuneSeparators}[1]{} 

\input{arm.fd}

\newcommand{\smallhagal}{{{\textarm{h}}}\!\!\!\!\!\!{\raisebox{0.5\depth}{\phantom|$^{\bullet}_{\pm}$}}}

\newcommand{\smallhagalplus}{{{\textarm{h}}}\!\!\!\!\!\!{\raisebox{0.5\depth}{\phantom|$^{\bullet}_{+}$}}}
\newcommand{\smallhagalminus}{{{\textarm{h}}}\!\!\!\!\!\!{\raisebox{0.5\depth}{\phantom|$^{\bullet}_{-}$}}}

\newcommand{\smallhagalbare}{{\textarm{h}}}

\newcommand{\overbar}[1]{\mkern 1.5mu\overline{\mkern-1.5mu#1\mkern-1.5mu}\mkern 1.5mu}

\newcommand{\del}[0]{\partial}

\let\baraccent=\=
\renewcommand{\=}[1]{\stackrel{#1}{=}}
\newcommand{\id}[0]{\mathbb{I}}

\DeclareSymbolFontAlphabet{\mathbb}{AMSb}

\begin{document}
	
	\pagestyle{plain}

	\makeatletter
	\@addtoreset{equation}{section}
	\makeatother
	\renewcommand{\theequation}{\thesection.\arabic{equation}}
	\pagestyle{empty}

	{\hfill MIT-CTP/5388}
	
	\vspace{1.5cm}
	
	\begin{center}
		
		{\LARGE \bf{Superpotentials from Singular Divisors}
			\\[10mm]}
	\end{center}

	\begin{center}
		\scalebox{0.95}[0.95]{{\fontsize{14}{30}\selectfont Naomi Gendler,$^{a}$ Manki Kim,$^{b}$ Liam McAllister,$^{a}$}} \vspace{0.35cm}
		\scalebox{0.95}[0.95]{{\fontsize{14}{30}\selectfont Jakob Moritz,$^{a}$ and Mike Stillman$^c$}}
	\end{center}

	\begin{center}
		\vspace{0.25 cm}
		\textsl{$^{a}$Department of Physics, Cornell University, Ithaca, NY 14853, USA}\\
		\textsl{$^{b}$Center for Theoretical Physics, MIT, Cambridge, MA 02138, USA}\\
		\textsl{$^{c}$Department of Mathematics, Cornell University, Ithaca, NY 14853, USA}\\
		
		\vspace{1cm}
		\normalsize{\bf Abstract} \\[8mm]

	\end{center}
We study Euclidean D3-branes wrapping
divisors $D$ in Calabi-Yau orientifold compactifications of type IIB string theory.
Witten's counting of fermion zero modes in terms of the cohomology of the structure sheaf $\mathcal{O}_D$ applies when $D$ is smooth, but we argue that
effective divisors of Calabi-Yau threefolds typically have singularities along rational curves.
We generalize the counting of fermion zero modes to such singular divisors, in terms of the cohomology of the structure sheaf $\mathcal{O}_{\overbar{D}}$ of the \emph{normalization} $\overbar{D}$ of $D$.
We establish this by detailing compactifications in which the  singularities can be unwound by passing through flop transitions, giving a physical incarnation of the normalization process. Analytically continuing the superpotential through the flops, we find that singular divisors whose normalizations are rigid can contribute to the superpotential: specifically,
$h^{\bullet}_{+}(\mathcal{O}_{\overbar{D}})=(1,0,0)$ and $h^{\bullet}_{-}(\mathcal{O}_{\overbar{D}})=(0,0,0)$
give a sufficient condition for a contribution.
The examples that we present feature infinitely many isomorphic geometric phases, with corresponding
infinite-order monodromy groups $\Gamma$.  We use the action of $\Gamma$ on effective divisors to determine the exact effective cones, which have infinitely many generators.
The resulting nonperturbative superpotentials are
Jacobi theta functions,
whose modular symmetries suggest the existence of strong-weak coupling dualities involving inversion of divisor volumes.

	\begin{center}
		\begin{minipage}[h]{15.0cm}

		\end{minipage}
	\end{center}
	\newpage
	\setcounter{page}{1}
	\pagestyle{plain}
	\renewcommand{\thefootnote}{\arabic{footnote}}
	\setcounter{footnote}{0}
	%
	%
	\tableofcontents
	\newpage

	\section{Introduction}\label{sec:intro}

The vacuum structure of a four-dimensional $\mathcal{N}=1$ supersymmetric
string compactification is encoded in its K\"ahler potential and superpotential.
The superpotential is holomorphic and
strongly constrained by non-renormalization theorems.
In particular, in type IIB string theory compactified on an O3/O7 orientifold $X$ of a Calabi-Yau (CY) threefold containing three-form flux \cite{Giddings:2001yu}, the flux superpotential \cite{Gukov:1999ya} is independent of the K\"ahler moduli of $X$, to all perturbative orders in the string loop and $\alpha'$ expansions \cite{Dine:1986vd}.
Thus, the potential for the K\"ahler moduli is governed, in many parameter regimes, by nonperturbative corrections to the superpotential, particularly from Euclidean D3-branes wrapped on holomorphic four-cycles, i.e.~effective divisors,
in $X$ \cite{Witten:1996bn}.
Characterizing such nonperturbative superpotential terms is a step toward understanding the landscape of low energy theories that arise in flux compactifications.

Evaluating the nonperturbative superpotential involves computing the D3-brane partition function for every effective divisor that can be wrapped by a Euclidean D3-brane.
Each effective divisor $D$ can in principle contribute a term
	\begin{equation}\label{eq:ED3superpotential}
		W\supset \mathcal{A}_D(z,\tau)e^{-2\pi \text{Vol}(D)-2\pi i \int_D C_4}\, ,
	\end{equation}
	where $\text{Vol}(D)$ is the volume of $D$
and $C_4$ is the Ramond-Ramond
four-form in ten dimensions. The Pfaffian $\mathcal{A}_D(z,\tau)$ can in general depend on all moduli except the K\"ahler moduli, and we have made explicit the dependence on the axiodilaton $\tau$ and the complex structure moduli $z$ of $X$.

Computing the numerical value of the Pfaffian at a given point in moduli space is often difficult \cite{Witten:1996hc,Ganor:1996pe,Buchbinder:2002pr,Cvetic:2012ts,Kerstan:2012cy,Blumenhagen:2006xt}.\footnote{For a review of this subject, see \cite{Blumenhagen:2009qh}.}
However, if the instanton solution has exact fermion zero modes beyond the two universal ones then $\mathcal{A}_D$ is identically zero.
The counting of fermion zero modes relies on data that is essentially topological, and is more readily computed than $\mathcal{A}_D(z,\tau)$ itself.  Thus, in favorable circumstances, one can
enumerate classes of effective divisors $D$ that support Euclidean D3-branes whose Pfaffians $\mathcal{A}_D(z,\tau)$ are \emph{not identically zero}.  This serves as a first step toward characterizing the nonperturbative superpotential.

For smooth divisors, and in the absence of fluxes,  the fermion zero modes are counted by the dimensions of the orientifold-graded
sheaf cohomology groups $H^i_{\pm}(D,\mathcal{O}_D)$.
Divisors $D$ that are rigid, connected, and orientifold-even, i.e.~with \cite{Witten:1996bn,Grassi:1997mr}
\begin{equation}\label{eq:Witten96}
		h^\bullet_+(D,\mathcal{O}_D):=\dim{H^\bullet_{+}(D,\mathcal{O}_D)}=(1,0,0)\, ,\quad h^\bullet_-(D,\mathcal{O}_D):=\dim{H^\bullet_{-}(D,\mathcal{O}_D)}=0\, ,
	\end{equation}
and that --- we reiterate --- are also \emph{smooth},
have Pfaffians that are not identically zero, and so generically contribute to the superpotential.

\begin{figure}[h!]
\centering
\includegraphics[scale=3.5]{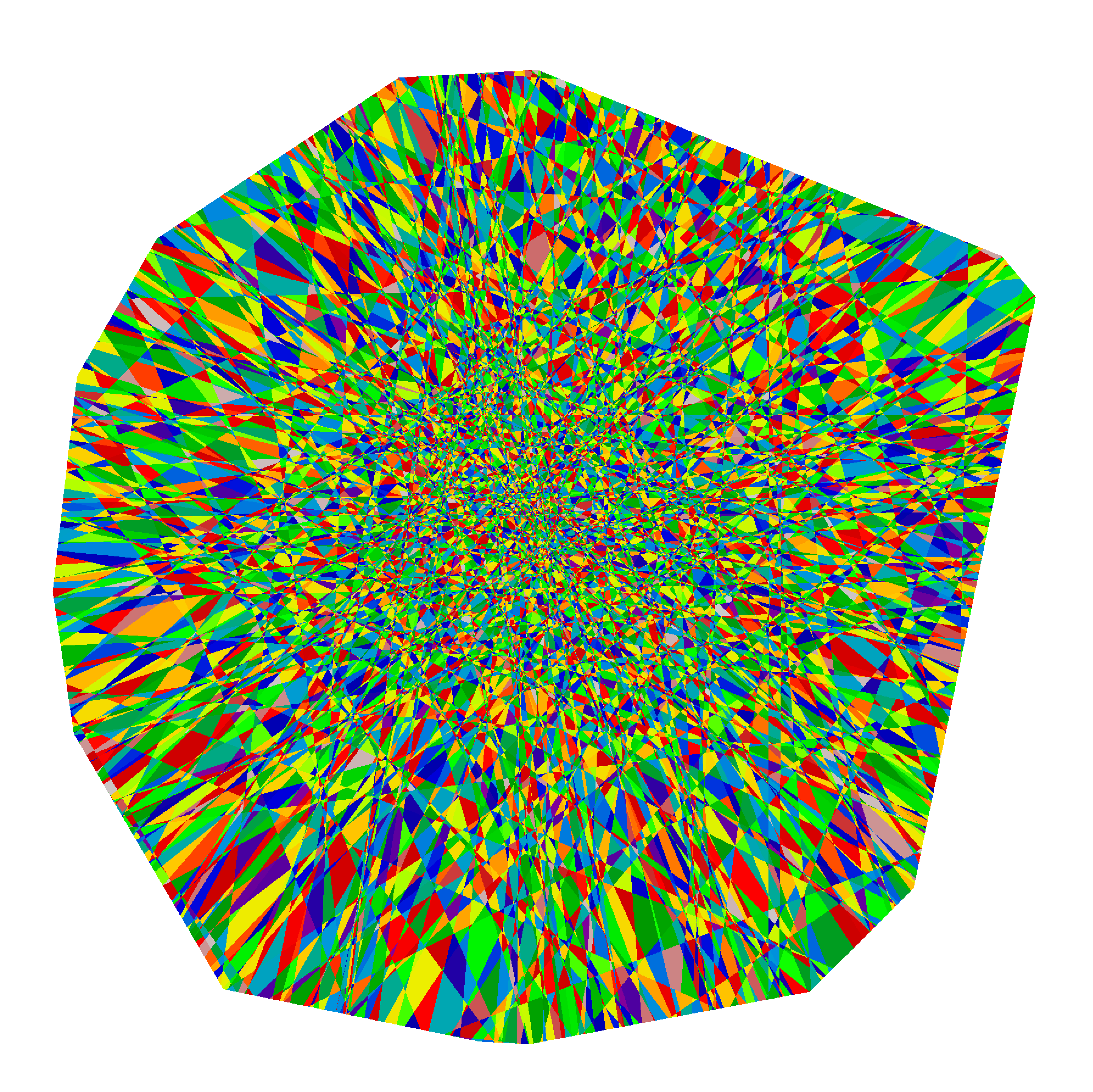}
\caption{A cross-section of the secondary fan of fine, regular, star triangulations of the largest polytope in the Kreuzer-Skarke database (with $h^{1,1}=491)$, with neighboring triangulations corresponding to the same Calabi-Yau combined. Each interior boundary corresponds to a bistellar flip, which induces star-crossing singularities on divisors. Figure courtesy of Andres Rios-Tascon.}
\label{491_fan}
\end{figure}

Unfortunately, the applicability of the above results is severely limited: an effective divisor class in a general Calabi-Yau threefold often has no smooth holomorphic representative.
Rather, as we will show in \S\ref{sec:flop}, effective divisors in Calabi-Yau threefolds
typically have \emph{singularities along rational curves}. The prevalence of such singularities is linked to the prevalence of flops. Specifically, in a given Calabi-Yau hypersurface in a toric variety $V$ obtained from triangulating a reflexive polytope, bistellar flips of the triangulation often induce flops of the Calabi-Yau threefold.\footnote{Only if the singularity in the ambient variety associated to the flip does not intersect the Calabi-Yau is no flop induced.}
The number of inequivalent triangulations grows rapidly with $h^{1,1}(V)$ \cite{Demirtas:2020dbm}: see Figure \ref{491_fan}. Flops are likewise common in complete intersection Calabi-Yau threefolds: see \cite{Brodie:2021toe} for a recent investigation. In this sense, flops, and correspondingly divisor singularities along rational curves, appear to be ubiquitous, at least in hypersurface and complete intersection Calabi-Yau threefolds.\footnote{We note that the extended K\"ahler cone is contained in the effective cone, so as a consequence of the abundance of flops the ample cone is typically much smaller than the effective cone. Thus, our claim that divisor singularities along rational curves are typical is not in contradiction with the well-known fact that (suitable multiples) of divisor classes in the ample cone have smooth generic representatives.}

For such singular $D$, satisfying \eqref{eq:Witten96} does not imply the existence of a nonvanishing contribution.
Thus, in order to understand the overall structure of the nonperturbative superpotential, one would very much like to adapt the result \eqref{eq:Witten96} to apply to divisors with singularities along rational curves. Doing so is the main purpose of the present work.
	
\begin{figure}
\centering
\begin{minipage}[t]{0.45\linewidth}
  \centering
  \includegraphics[width=\linewidth]{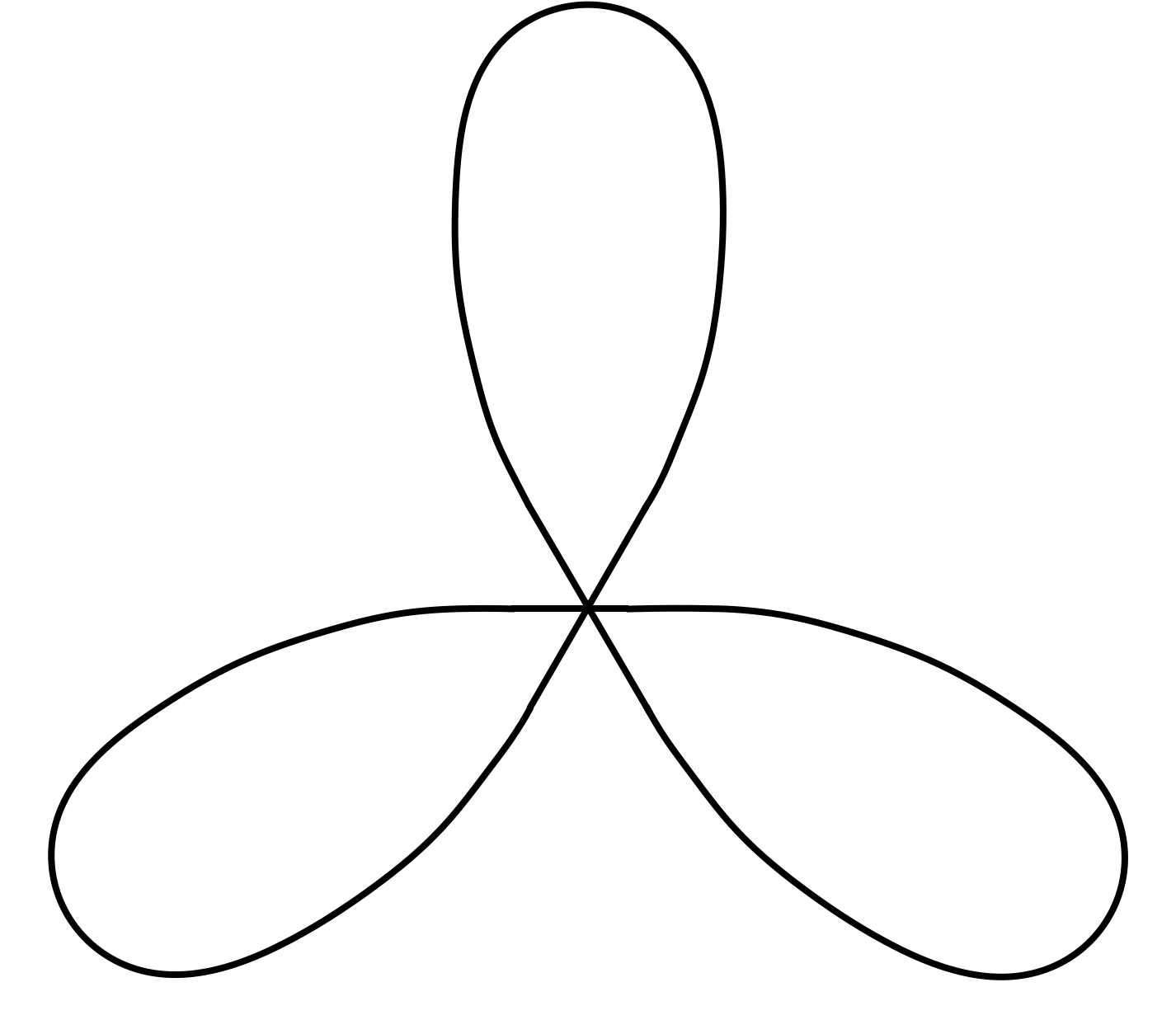}
  \caption{A star-crossing singularity with $k=3$ locally intersecting components.}
  \label{fig:normalcrossing}
\end{minipage}%
\hfil
\begin{minipage}[t]{.45\linewidth}
  \centering
  \includegraphics[width=\linewidth]{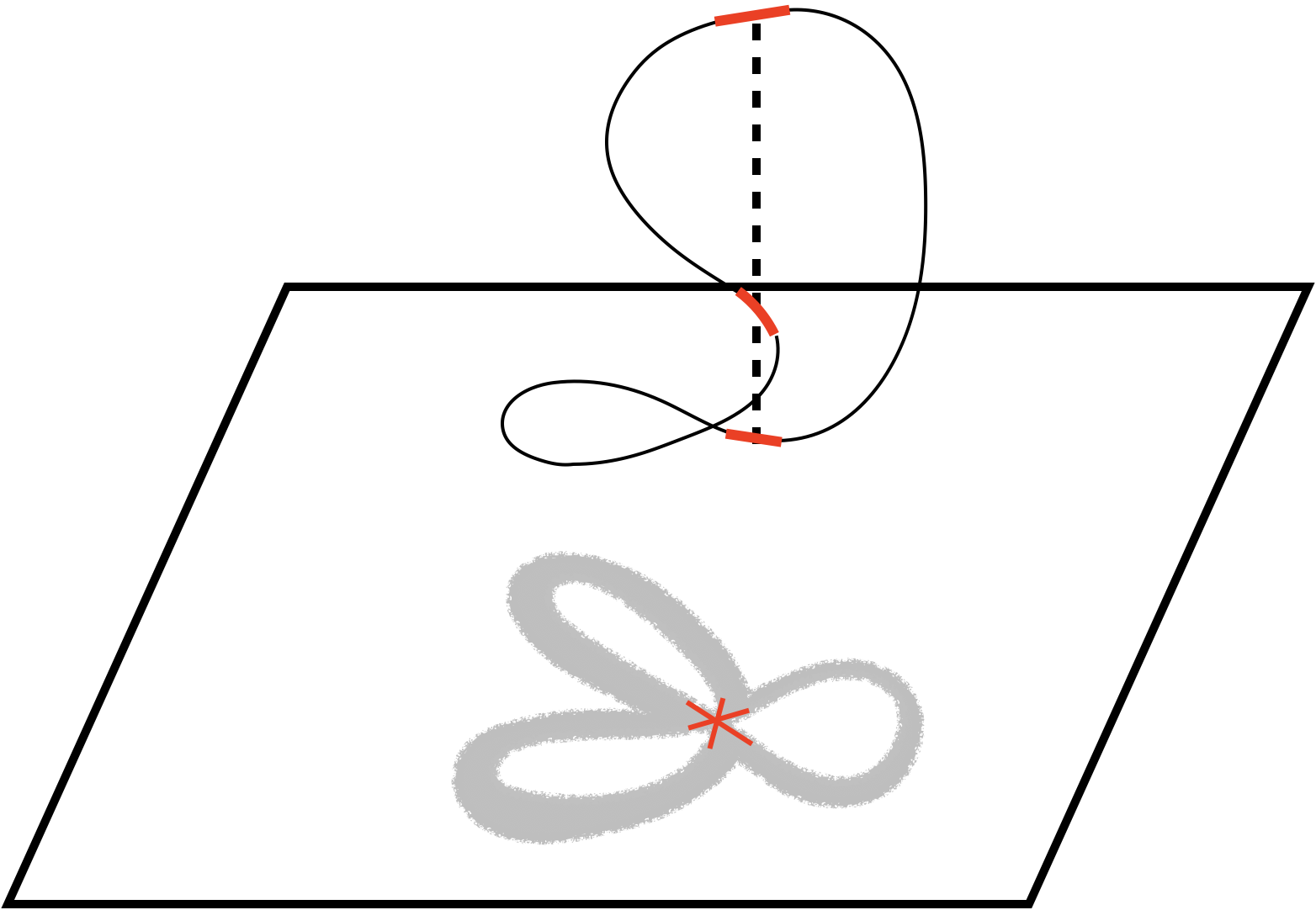}
  \caption{The normalization of a $k=3$ star-crossing singularity.}
  \label{fig:normalization}
\end{minipage}
\end{figure}

Specifically, we suppose that a divisor $D \subset X$ is singular along a rational curve $\mathcal{C} \subset D$,
and that in a local neighborhood around $\mathcal{C}$, $D$ takes the form of $k \ge 2$ smooth irreducible surfaces intersecting each other transversely along $\mathcal{C}$: see Figure~\ref{fig:normalcrossing}.
When $k=2$, such a singularity is a normal crossing singularity (see e.g.~\cite{Steenbrink}).
However, in general one encounters cases with arbitrarily large $k$, so we will speak of \emph{star-crossing singularities}, with the understanding that the local description is always as stated above.

Our main claim is that for divisors with star-crossing singularities along smooth rational curves,
fermion zero modes are counted by replacing $D$ with $\overbar{D}$ in \eqref{eq:Witten96}, with $\overbar{D}$ the \emph{normalization} of the singular divisor $D$.  Intuitively, the normalization $\overbar{D}$ is a (partial) resolution of $D$ resulting from splitting apart the components that intersect along the curve $\mathcal{C}$: see Figure~\ref{fig:normalization}. A precise definition of the normalization $\overbar{D}$ is given in \S\ref{sec:defnorm}.

We claim that for $D$ a divisor with star-crossing singularities along smooth rational curves, if
\begin{equation}\label{eq:newboxtimes}
		h^\bullet_+(\overbar{D},\mathcal{O}_{\overbar{D}})=(1,0,0)\, ,\quad h^\bullet_-(\overbar{D},\mathcal{O}_{\overbar{D}})=0\, ,
	\end{equation}
then there is a nonperturbative superpotential term from Euclidean D3-branes, even if $D$ does not fulfill \eqref{eq:Witten96}.
We will use the notation\footnote{The symbol $\smallhagalbare$ is the rune `hagal', which denotes the sound $h$, and, conveniently, depicts a star-crossing singularity with $k=3$.}
\begin{equation}\label{eq:boxtimes}
{\smallhagal}(D) := h^\bullet_{\pm}\left(\overbar{D},\mathcal{O}_{\overbar{D}}\right)\,,
\end{equation}
in terms of which the condition \eqref{eq:newboxtimes} takes the form
\begin{equation}\label{eq:hagalboxtimes}
		{\smallhagalplus}(D)=(1,0,0)\, ,\quad {\smallhagalminus}(D)=0\, .
	\end{equation}
We refer to divisors satisfying \eqref{eq:hagalboxtimes} as \textit{star-rigid}.

We will arrive at the criterion \eqref{eq:hagalboxtimes}
by examining Calabi-Yau compactifications for which the normalization map acquires a physical incarnation as the birational map relating two geometric phases that are connected by a flop transition.\footnote{Strictly speaking, the flop transition and the normalization map differ by the blowups of one or more smooth points, but this does not affect the counting of fermion zero modes: see \S\ref{sec:defnorm}.}
Concretely, we suppose that $D$ is
a divisor with star-crossing singularities in a Calabi-Yau threefold $X$ that is birationally equivalent to another Calabi-Yau threefold $X'\sim X$ in which the corresponding divisor $D'$ is smooth.  Such a
pair of birationally-equivalent Calabi-Yau threefolds is connected
by a series of flop transitions, and the fermion zero modes of half-BPS instantons can be tracked by analytic continuation from one phase to the other without encountering discontinuities. Fermion zero modes can thus be counted in a suitable phase where the corresponding divisor is smooth and \eqref{eq:Witten96} applies. Using a local description of flop transitions (see \S\ref{sec:flop}), we find that  ${\smallhagalbare}(D)$   is a birational invariant, and coincides with $h^\bullet(D,\mathcal{O}_D)$ when $D$ is smooth, and must therefore count zero modes in every phase.

In \S\ref{sec:examples} we will detail examples for which the above argument applies directly, i.e.~where the star-crossing singularities of a divisor $D$ can be unwound by suitably transitioning between adjoining geometric phases using flop transitions.
However, we also conjecture that $\smallhagal$ counts zero modes more generally, for any divisor $D$ with star-crossing singularities along rational\footnote{Divisors with singularities along higher-genus curves will generally support additional zero modes from strings stretching from one component to another: see \S\ref{subsec:openstringargument}.} curves.

The two
geometries we use to illustrate the counting of fermion zero modes turn out to have striking properties, which we explore in the second part of this work (\S\ref{sec:examples}).
Both feature infinite series of flops connecting isomorphic geometric phases, with corresponding infinite-order monodromy groups $\Gamma$ that act on curve and divisor classes.  By examining the action of $\Gamma$ on effective divisors, we determine the exact effective cones in these geometries.  These cones have infinitely many generators related by the action of $\Gamma$, and give rise to infinitely many nonperturbative superpotential terms.  Summing up these contributions, we express the nonperturbative superpotential in each geometry in terms of a
Jacobi theta function.  The modular transformation properties of the theta function suggest the existence of a connection between two regimes of the effective theory related by inversion of the volumes of certain divisors.  We speculate that these configurations are related by a heretofore unknown strong-weak coupling duality.

The organization of this paper is as follows.  In \S\ref{sec:cone} we review general facts about flops and about nonperturbative superpotential terms from Euclidean D3-branes.
In \S\ref{sec:flop} we recall the local description of a flop, we explain how flops produce star-crossing singularities in divisors, and we give a precise definition of the normalization of a divisor.
In \S\ref{sec:analytic} we analytically continue the superpotential through a series of flops that connect a divisor with a star-crossing singularity to its normalization.  We present two explicit examples in \S\ref{sec:examples}.  In each case we compute the exact effective cone and express the nonperturbative superpotential in terms of a Jacobi theta function.
We conclude in \S\ref{sec:conclusions}.

\section{Cones, flops, and superpotentials}\label{sec:cone}

To fix our terminology and notation we will review some well known aspects of Calabi-Yau geometry, with an emphasis on flop transitions, as well as Calabi-Yau compactifications of type IIB string theory.

\subsection{Mori, K\"ahler, and effective cones}
Let $X$ be a  Calabi-Yau threefold, and let $\mathcal{K}_X\subset H^2(X,\mathbb{R})$ be its \textit{K\"ahler cone}, i.e.~the set of cohomology classes of closed $(1,1)$-forms $J=-2ig_{a\bar{b}}dz^a\wedge d\bar{z}^{\bar{b}}$ such that $g_{a\bar{b}}$ defines a smooth Hermitian metric on $X$. Here the holomorphic indices $a,b$ range from $1$ to $3$, and the antiholomorphic indices $\bar{a},\bar{b}$ range from $\bar{1}$ to $\bar{3}$.
Via Yau's proof of the Calabi conjecture, for each choice $J\in \mathcal{K}_X$ there exists a unique Ricci-flat metric $g_{a\bar{b}}$ on $X$.
Integer classes in $\mathcal{K}_X$ define ample line bundles, which are in one to one correspondence with ample divisor classes. The cone generated by line bundles that have non-trivial global holomorphic sections is the \textit{effective cone} $\mathcal{E}_X$, and integer classes in $\mathcal{E}_X$ define \textit{effective divisor} classes. We have $\mathcal{K}_X\subset \mathcal{E}_X$. The closure $\overline{\mathcal{E}_X}$ is the \textit{pseudo-effective cone}.

The cone dual to $\mathcal{K}_X$, denoted $\mathcal{M}_X\subset H^4(X,\mathbb{R})$, is the \textit{Mori cone}, and integer classes in $\mathcal{M}_X$ define \textit{effective curve} classes. The cone dual to $\mathcal{E}_X$, denoted $\text{Mov}(X)$, is the \textit{cone of movable curves},
and we have $\text{Mov}(X)\subset \mathcal{M}_X$.
Effective curves $[\mathcal{C}]\in \mathcal{M}_X$, effective divisors $[D]\in \mathcal{E}_X$, and the Calabi-Yau $X$  are calibrated with respect to the K\"ahler form $J\in \mathcal{K}_X$,
i.e.~their volumes measured in the metric $g_{a\bar{b}}$ are given by
\begin{equation}
	\text{Vol}(\mathcal{C})=\int_{\mathcal{C}}J\, ,\quad \text{Vol}(D)=\int_{D}\frac{1}{2}J\wedge J\, ,\quad \mathcal{V}:=\text{Vol}(X)=\int_X \frac{1}{6}J\wedge J\wedge J\, .
\end{equation}
A few words are in order about how the cones defined above can be computed in concrete examples. We will focus on smooth Calabi-Yau hypersurfaces $X$ in toric fourfolds $V$ obtained from the Kreuzer-Skarke dataset \cite{Kreuzer:2000xy}, and for simplicity we further restrict to favorable embeddings $X\subset V$, i.e.~cases in which $h^{1,1}(V)=h^{1,1}(X)$. As every effective curve in $\mathcal{M}_X$ is also effective viewed as a curve in $V$, we have the natural inclusion $\mathcal{M}_X\subset \mathcal{M}_V$, where $\mathcal{M}_V$ is the Mori cone of $V$. The Mori cone of a toric $d$-fold is generated by the torus-invariant subvarieties associated with the $(d-1)$-dimensional cones of the toric fan, so obtaining $\mathcal{M}_V$ is computationally straightforward. It is often the case that a set of birationally-equivalent but non-isomorphic toric varieties $\{V_i\}$ all have the same Calabi-Yau threefold $X$ as their generic anticanonical hypersurface. This happens when the $V_i$ are related to each other via bistellar flip transitions, in such a way that $X$ does not intersect the singularities of the ambient varieties that occur along the transition loci. Using this, we may obtain a better approximation of $\mathcal{M}_X$ via $\mathcal{M}_X\subset \mathcal{M}^{\cap}_V:=\cap_i \mathcal{M}_{V_i}$.

One can use $\mathcal{M}^\cap_V$ as a starting point to obtain the exact Mori cone of $X$ \cite{WGV}. For this, one uses the fact that $\mathcal{M}_X$ is the BPS cone of M-theory compactified on $X$. Using tools developed in \cite{Hosono:1993qy,Hosono:1994ax} one can compute the BPS indices --- Gopakumar-Vafa (GV) invariants --- associated with integer points in $\mathcal{M}^\cap_V$.\footnote{For progress in computing GV invariants at large $h^{1,1}(X)$, see the upcoming work \cite{compmirror}.} We denote by $\mathcal{M}_X^{\mathrm{GV}}$ the cone over all sites in $H_2(X,\mathbb{Z})\simeq H_2(V,\mathbb{Z})$ with non-vanishing GV invariants, and we can use this cone to further bound the Mori cone of $X$, $\mathcal{M}_X^{\mathrm{GV}}\subset \mathcal{M}_X\subset \mathcal{M}^{\cap}_V$. In many cases, including all examples in this paper, one finds\footnote{One can obtain the exact Mori cone $\mathcal{M}_X$ even when $\mathcal{M}_X^{\mathrm{GV}} \neq \mathcal{M}^{\cap}_V$ \cite{WGV}, but this will not be necessary in our examples. } $\mathcal{M}_X^{\mathrm{GV}}=\mathcal{M}^{\cap}_V$, which implies $\mathcal{M}_X=\mathcal{M}_X^{\mathrm{GV}}=\mathcal{M}^{\cap}_V$.  The K\"ahler cone is then obtained by computing the dual cone of $\mathcal{M}_X$.

The effective cone of the ambient variety is easily computed: it is generated by the torus-invariant divisors associated with the $h^{1,1}(V)+4$ edges of the toric fan. In general, one has a map $H^2(V,\mathbb{Z})\rightarrow H^2(X,\mathbb{Z})$ defined via intersecting a divisor class of $V$ with the Calabi-Yau hypersurface $X$, and for favorable embeddings this is an isomorphism. In particular, every effective divisor $\widehat{D}\subset V$ defines an effective divisor $D:=\widehat{D}\cap X$ in $H^2(X,\mathbb{Z})$, and thus $\mathcal{E}_V\subset \mathcal{E}_X$.\footnote{Note that the inclusion is reversed in comparison to the inclusion of Mori cones.} In general $\mathcal{E}_V \subsetneq  \mathcal{E}_X$, and we will call a divisor class $[D]\in (\mathcal{E}_X\backslash \mathcal{E}_V)\cap H^2(X,\mathbb{Z})$ an \textit{autochthonous divisor}. We will comment on how to compute $\mathcal{E}_X$ in \S\ref{sec:flop_review}.

\subsection{Flop transitions and the cone of effective divisors}  \label{sec:flop_review}
For each point on the boundary $\del \overline{\mathcal{K}_X}$ of the (closure of the) K\"ahler cone, a set of effective curves shrinks to zero volume, and thus the Calabi-Yau metric degenerates. For a generic choice of $J\in \del \overline{\mathcal{K}_{X}}$, all vanishing curves are integer multiples of an irreducible curve class $\mathcal{C}_v$, and we distinguish among the following possibilities that can occur at finite distance in moduli space (see e.g. \cite{Witten:1996qb}):
\begin{enumerate}
	\item $\mathcal{C}_v$ shrinks, but no effective divisors shrink. We call this a \textit{flop wall} of $\mathcal{K}_X$.
	\item  $\mathcal{C}_v$ shrinks, and an effective divisor $D_v$ containing $\mathcal{C}_v$ degenerates to a curve of genus $g$, leading to non-abelian enhancement to $\mathfrak{su}(2)$ \cite{Aspinwall:1995xy,Katz:1996ht}.
	\item  $\mathcal{C}_v$ shrinks, and an effective divisor $D_v$ containing $\mathcal{C}_v$ shrinks to a point. This is a strong coupling boundary, featuring tensionless strings.
\end{enumerate}
Throughout this paper, we will focus on (a), flop walls. Importantly, it makes sense to continue the Calabi-Yau geometry past a flop wall into a region where $\int_{\mathcal{C}_v}J<0$, producing a new Calabi-Yau threefold $X'$. In $X'$ the class $[\mathcal{C}_v]$ ceases to be effective, but instead $-[\mathcal{C}_v]$ becomes effective, and the threefolds $X$ and $X'$ are birationally equivalent, though not generally isomorphic.
Indeed, a Calabi-Yau threefold is, via Wall's theorem \cite{Wall}, uniquely specified as a real manifold by its Hodge numbers $(h^{1,1},h^{2,1})$, triple intersection numbers, and second Chern class. While the Hodge numbers are invariant across a flop transition, the triple intersection numbers and second Chern class transform non-trivially \cite{mcduff1994j,wilson1999flops},
\begin{align}\label{eq:tripleintC2_trafo}
	\int_X D\wedge D'\wedge D''&\longrightarrow \int_{X'} D\wedge D'\wedge D''=\int_X D\wedge D'\wedge D''-n_{\mathcal{C}}^0 (\mathcal{C}\cdot D)(\mathcal{C}\cdot D')(\mathcal{C}\cdot D'')\, ,\nonumber\\
	c_2(X)&\longrightarrow c_2(X')=c_2(X)+2n_{\mathcal{C}}^0\, [\mathcal{C}]\, ,
\end{align}
where $D$, $D'$ and $D''$ are divisor classes in $H^2(X)$, $\mathcal{C}\cdot D$ denotes the intersection pairing between the vanishing curve and $D$, and $n_{\mathcal{C}}^0$ denotes the genus zero GV invariant of the class $[\mathcal{C}]$, which for a flop transition is equal to the number of holomorphic representatives in the class $[\mathcal{C}]$.

For later reference, we also write down how the holomorphic Euler characteristic $\chi(D,\mathcal{O}_D)$ of a divisor $D$ transforms:
\begin{equation}\label{eq:chihol_trafo}
	\chi(D,\mathcal{O}_D)\longrightarrow \chi(D,\mathcal{O}_D)+n_{\mathcal{C}}^0 \frac{( \mathcal{C}\cdot D+1) ( \mathcal{C} \cdot D) ( \mathcal{C}\cdot D-1)}{3!}\, ,
\end{equation}
and note that $\chi(D,\mathcal{O}_D)$ remains invariant if $ \mathcal{C} \cdot D\in \{-1,0,1\}$.

Flops in threefolds have been classified in \cite{KatzMorrison}. Those associated to codimension-one walls of the K\"ahler cones of smooth Calabi-Yau threefolds correspond to the shrinking of rational curves with normal bundle $\mathcal{O}(-1)\oplus\mathcal{O}(-1)$, $\mathcal{O}\oplus\mathcal{O}(-2)$, or $\mathcal{O}(1)\oplus\mathcal{O}(-3)$ \cite{Laufer}. In the simplest case with normal bundle $\mathcal{O}(-1)\oplus\mathcal{O}(-1)$ --- which will be the relevant case in this paper --- the local neighborhood around the vanishing curve is well-approximated by the total space of its normal bundle, the \textit{resolved conifold} \cite{atiyah1958analytic,Candelas:1989js}. Examples of flops with normal bundle $\mathcal{O}\oplus\mathcal{O}(-2)$ and $\mathcal{O}(1)\oplus\mathcal{O}(-3)$ appear in \cite{Laufer,Pinkham1983}.

The union $\mathcal{K}$ of all the K\"ahler cones of a set of Calabi-Yau threefolds related to each other by flop transitions is called the \textit{extended K\"ahler cone}. By construction, exterior walls of $\mathcal{K}$ are either of type (b) or (c):  at each wall of $\mathcal{K}$ an effective divisor shrinks. Thus, given the extended K\"ahler cone $\mathcal{K}$, one can determine the generators of the effective cone by enumerating the divisor classes that shrink on $\del \overline{\mathcal{K}}$.
A related result is that the image of the extended K\"ahler cone under the map $J\mapsto \frac{1}{2}J\wedge J\in H^4(X,\mathbb{R})$ is equivalent to the cone of movable curves $\text{Mov}(X)$, i.e.~to the dual of the effective cone \cite{Alim:2021vhs}.

In our examples in \S\ref{sec:examples} we will determine the exact extended K\"ahler cone $\mathcal{K}$ by enumerating all possible geometrical phases related to each other via flop transitions, and use this to determine the effective cone $\mathcal{E}_X$.
	
\subsection{Symmetric flops}\label{sec:symm_flops}
Though the formulas \eqref{eq:tripleintC2_trafo} suggest that generic pairs of Calabi-Yau threefolds connected via flops are not isomorphic to each other, in special circumstances there may exist a change of basis of $H^2(X,\mathbb{Z})$,
\begin{equation}
	\{H_a\}_{a=1}^{h^{1,1}(X)}\mapsto \{H'_a\}_{a=1}^{h^{1,1}(X)}\, ,\quad H'_a={\Lambda_a}^b H_a\, ,
\end{equation}
parameterized by an integer matrix $\Lambda\in GL(h^{1,1}(X),\mathbb{Z})$ acting on the basis divisors, such that
\begin{equation}
	\int_X H_a \wedge H_b \wedge H_c=\int_{X'} H'_a \wedge H'_b \wedge H'_c\, ,\quad \text{and} \quad \int_X H_a\wedge c_2(X)=\int_{X'} H'_a\wedge c_2(X')\, .
\end{equation}
In this case $X\simeq X'$ via Wall's theorem \cite{Wall}, and we follow \cite{Brodie:2021ain,Brodie:2021toe} in calling such a flop a \emph{symmetric flop}. Even though the Calabi-Yau threefolds on the two sides of the flop are isomorphic, upon continuing through the flop wall the change of basis $\Lambda$ acts non-trivially on the curve and divisor classes. Thus, a path in K\"ahler moduli space starting and ending at suitable points in geometric phases that are related to each other via a symmetric flop gives rise to a non-trivial monodromy in K\"ahler moduli space. If $\Lambda^\ell \neq \id$ for all nonzero $\ell \in \mathbb{Z}$, it follows that there are in fact infinitely many isomorphic phases related to each other via symmetric flops, as in \cite{Brodie:2021ain,Brodie:2021toe}.  We will encounter such behavior in the examples of \S\ref{sec:examples}.

In such a case, the monodromy group $\Gamma$ contains the infinite-order subgroup $\Gamma_{\Lambda}:=\{\Lambda^\ell\, ,\, \ell\in \mathbb{Z}\}$. In our examples we will see that the action of $\Gamma_{\Lambda}$ can be used to show that the Euclidean D3-brane superpotentials can be summed into certain Jacobi forms. Moreover, we will be able to use the action of the monodromy group to obtain exact, infinitely-generated effective cones.
\subsection{Nonperturbative superpotentials in type IIB orientifolds}

We conclude this section with a brief review of the conditions for a
contribution to the superpotential in compactifications of type IIB string theory.

A prerequisite for an instanton to generate a nonzero superpotential contribution  is that the instanton has precisely the two universal zero modes corresponding to the measure of the superspace integral.

For instance, in M-theory on a compact Calabi-Yau fourfold $Y$, an M5-brane wrapping a holomorphic divisor $D\subset Y$ may contribute to the superpotential.
As shown in \cite{Witten:1996bn}, for \textit{smooth} divisors, and in the absence\footnote{In type IIB orientifolds, zero modes can be lifted by background three-form flux, or by a nontrivial gauge bundle on the D3-brane worldvolume, allowing a wider range of divisors $D$ to support superpotential terms \cite{Bianchi:2011qh,Bianchi:2012pn,Bianchi:2012kt,Grimm:2011dj}. Such effects will not be relevant for this work.  Furthermore, we note that zero mode counting for bound states of irreducible Euclidean D3-branes can be subtle: some of the zero modes of one component can be lifted by the presence of the other component, potentially leading to non-trivial \textit{poly-instanton} corrections \cite{Blumenhagen:2012kz}. In this paper we will consider only irreducible Euclidean D3-branes.}
 of bulk fluxes, the zero modes are counted by the dimensions of the cohomology groups of the structure sheaf $\mathcal{O}_D$, $h^{i}(D,\mathcal{O}_D) := \text{dim} \, H^{i}(D, \mathcal{O}_D)$. In particular, a sufficient condition for such an M5-brane to generate a superpotential term is that these dimensions obey
\begin{align}\label{eq:Witten96Fourfold}
h^{0}(D,\mathcal{O}_D)=1\, ,\quad  h^{i}(D,\mathcal{O}_D)=0 \quad \forall \, i>0\, ,
\end{align}
which we abbreviate as $h^{\bullet}(D,\mathcal{O}_D)=(1,0,0,0)$. Divisors satisfying this condition are referred to as \textit{rigid}, because they do not admit any massless bosonic deformations.

In $\mathcal{N}=1$ compactifications of type IIB string theory on Calabi-Yau orientifolds, the relevant cohomology groups are those that are even under the orientifold action, and a sufficient condition for a non-vanishing superpotential contribution from a Euclidean D3-brane on a smooth divisor $D$ is
\begin{equation}\label{eq:Witten96Threefold}
		h^\bullet_+(D,\mathcal{O}_{D})=(1,0,0)\, ,\quad h^\bullet_-(D,\mathcal{O}_{D})=0\, ,
	\end{equation}
where the bullet now stands in for $i=0, \, 1, \, 2$.

Witten's argument starts with the fact that given an M5-brane wrapped on a smooth divisor $D\subset Y$, its worldvolume fermions take values in the spinor bundle constructed out of the normal bundle of $D$ in $Y$.  One thus expects the condition \eqref{eq:Witten96Fourfold}, and likewise \eqref{eq:Witten96Threefold}, to be modified when the embedding of the divisor $D\subset Y$ is sufficiently singular so that the notion of a normal bundle becomes ambiguous.\footnote{The conormal bundle enjoys an unambiguous generalization to the conormal sheaf, which can be used as a starting point that leads to results consistent with those of the present work.}

The focus of this work is to provide a new prescription for divisors with singularities along rational curves, which --- as explained in the Introduction \S\ref{sec:intro} --- is the generic situation.\footnote{For a related discussion see e.g. \cite{Donagi:2010pd}.}

\section{Flop transitions and divisor singularities}\label{sec:flop}

We will now give an explicit local description of the simplest kind of flop transition, involving a local conifold geometry, and use this to define and understand star-crossing singularities.

\subsection{Local conifold geometry of a flop}

In \S\ref{sec:flop_review}, we reviewed how flop transitions relate distinct Calabi-Yau threefolds via curves that shrink to codimension-three singularities, at walls in the K\"ahler cone where no divisors shrink.
In this section, we will review local descriptions of the ensuing singular varieties and their small resolutions. We will use this to demonstrate that divisors intersecting the shrinking curves multiple times undergo non-trivial topological transitions, whereby codimension-one singularities are created.

As mentioned in \S\ref{sec:flop_review}, the possible normal bundles of rational curves that can shrink along flop walls are $\mathcal{O}(-1) \oplus \mathcal{O}(-1)$, $\mathcal{O} \oplus \mathcal{O}(-2)$, or $\mathcal{O}(1) \oplus \mathcal{O}(-3)$ \cite{Laufer,Pinkham1983,KatzMorrison}. In this work, we will focus on the simplest case of $\mathcal{O}(-1) \oplus \mathcal{O}(-1)$. Our discussion naturally extends to the other cases \cite{StillmanFlops}.

For a rational curve $\mathcal{C}\subset X$ with normal bundle $\mathcal{N}_{\mathcal{C}|X}=\mathcal{O}(-1)\oplus \mathcal{O}(-1)$ the local geometry around $\mathcal{C}\subset X$ is approximated by the resolved conifold $X^{\text{cf}}_{\text{res}}$ \cite{atiyah1958analytic,Candelas:1989js}. We can describe $X^{\text{cf}}_{\text{res}}$ locally as a codimension-two hypersurface in $\mathbb{C}^4 \times \mathbb{P}^1$ via
\begin{equation}\label{eq:resolution1}
	M\cdot \begin{pmatrix}
		\alpha\\
		\beta
	\end{pmatrix}=0\, ,\quad \text{with}\quad  M:=\begin{pmatrix}
		x & v\\
		u & y
	\end{pmatrix}\, ,
\end{equation}
with $(x,y,u,v)\in \mathbb{C}^4$ and $[\alpha:\beta]\in \mathbb{P}^1$. Note that \eqref{eq:resolution1} implies the constraint $\det M=0$, as $\alpha$ and $\beta$ cannot vanish simultaneously. Away from the locus $M=0$ we can solve for $[\alpha:\beta]$ in terms of $(x,y,u,v)$ and therefore we can define a blow-down map
\begin{equation}
	\phi_{\text{blow-down}}: \,X^{\text{cf}}_{\text{res}}\twoheadrightarrow X^{\text{cf}}_{\text{sing}}:=\{\det M=0\}\subset \mathbb{C}^4\, ,
\end{equation}
defined by ``forgetting" the coordinates $[\alpha:\beta]$.
The map $\phi_{\text{blow-down}}$ is one-to-one everywhere except at the origin $p_0:=\{M=0\}\in X^{\text{cf}}_{\text{sing}}$, and $\phi_{\text{blow-down}}^{-1}(p_0)\simeq \mathbb{P}^1$ defines an exceptional $\mathbb{P}^1$ that resolves the singularity at the origin of $X_{\text{sing}}^{\text{cf}}$, and is identical to the rational curve $\mathcal{C}$.

Crucially, the singular variety $X_{\text{sing}}^{\text{cf}}$ can alternatively be resolved by blowing up a different exceptional $\mathbb{P}^1$: we instead consider
\begin{equation}\label{eq:resolution2}
	M^T \cdot\begin{pmatrix}
		\tilde{\alpha}\\
		\tilde{\beta}
	\end{pmatrix}=0\, ,
\end{equation}
where the new exceptional $\tilde{\mathcal{C}}\simeq \mathbb{P}^1$ is likewise parameterized by homogeneous coordinates $[\tilde{\alpha}:\tilde{\beta}]\in \mathbb{P}^1$ over the origin of $\mathbb{C}^4$.  The two distinct resolutions are birationally equivalent, i.e.~there is a one-to-one map
\begin{equation}\label{eq:birationalmapX}
	X_{\text{res}}^{\text{cf}}\setminus\mathcal{C}\longleftrightarrow \widetilde{X}_{\text{res}}^{\text{cf}}\setminus\tilde{\mathcal{C}}\, ,
\end{equation}
and interpolating from one $X_{\text{res}}^{\text{cf}}$ to $\widetilde{X}_{\text{res}}^{\text{cf}}$ via shrinking $\mathcal{C}$ and then blowing up $\tilde{\mathcal{C}}$ is a flop transition.

For our purposes it will be important to understand how divisors that intersect flop curves behave under flop transitions. For simplicity we start our discussion with a divisor $D$ intersecting the curve $\mathcal{C}\simeq \mathbb{P}^1$ transversely in a single point. We can represent $D$ locally by the constraint $\alpha=0$, which by \eqref{eq:resolution1} implies $v=y=0$. Thus, the topology of $D$ is simply $\mathbb{C}^2$. Note however, that upon imposing the constraint $v=y=0$ the defining equation \eqref{eq:resolution1} collapses to $x\alpha=u\alpha=0$, which has two branches: one of them gives an isolated point $x=y=u=v=0$ and the other is our divisor $D$.

In the new phase, however, $v=y=0$ implies
\begin{equation}
	x\tilde{\alpha}+u\tilde{\beta}=0\, ,
\end{equation}
so in the flopped geometry, the divisor $D$ has the topology of $\mathbb{C}^2$ blown up at the origin.

We thus conclude that under a flop transition involving the shrinking of an exceptional curve $\mathcal{C}\simeq \mathbb{P}^1$, a divisor $D$ intersecting $\mathcal{C}$ transversely in a single point, i.e.~with intersection pairing $D\cdot \mathcal{C}=1$, gets blown up at the intersection point. The new exceptional curve $\tilde{\mathcal{C}}\simeq \mathbb{P}^1$ is contained in the new divisor $\widetilde{D}$, consistent with the intersection pairing $\widetilde{D}\cdot \tilde{\mathcal{C}}=-1$ that follows from the homology relations
\begin{equation}
	[\tilde{\mathcal{C}}]=-[\mathcal{C}]\, ,\quad [\widetilde{D}]=[D]\, .
\end{equation}
Conversely, a divisor $D$ that contains $\mathcal{C}$ with intersection pairing $D\cdot \mathcal{C}=-1$ will intersect the new effective curve $\tilde{\mathcal{C}}$ transversely after the flop transition. This result is well-known and appears, for instance, in the discussion of geometric transitions of orientifold planes in Calabi-Yau orientifold models \cite{Denef:2005mm,Carta:2020ohw}.

Next, we would like to understand what happens to a divisor $D$ that intersects the conifold curve $\mathcal{C}$ in $k>1$ points. Locally, we can represent such a divisor as the vanishing locus of a degree $k$ polynomial in the homogeneous coordinates $[\alpha:\beta]$,
\begin{equation}
	0=f_k=\sum_{i=0}^{k}c_i \alpha^{i}\beta^{k-i}\, .
\end{equation}
Generally, the coefficients $c_i$ can be functions of the coordinates $(x,y,u,v)$ but close to the conifold curve $\mathcal{C}$ we can treat them as constants, $c_i\rightarrow c_i|_{x=y=u=v=0}$. Therefore, again locally, the defining polynomial $f_k$ factorizes into $k$ solution branches,  each of which is equivalent to a divisor intersecting $\mathcal{C}$ once. Globally, $f_k$ need not factorize and the $k$ local branches may correspond to $k$ disjoint local neighborhoods of a single irreducible divisor.

But, by virtue of our discussion of the special case $k=1$, we already know how each of the $k$ distinct local branches transforms under the flop: each sheet locally looks like a copy of $\mathbb{C}^2$, and the flop induces a blowup transition in all sheets. As the blowup $\mathbb{P}^1$ of each sheet is identified with the exceptional curve $\tilde{\mathcal{C}}$ in the resolved conifold, the flop transition glues together all $k$ sheets by identifying their respective blowup $\mathbb{P}^1$'s. It is straightforward to show that for generic choices of the $c_i$, each pair in the set of $k$ sheets intersects each other transversely in the exceptional curve $\tilde{\mathcal{C}}$. For $k=2$ this means that the divisor $D$ has a \textit{normal crossing singularity} along the exceptional curve $\mathcal{C}$. For general $k$ we obtain a slightly more general divisor singularity of $k$ local sheets that intersect each other transversely pairwise.  As explained in the Introduction, we term this a star-crossing singularity.

Now suppose that, in some geometric phase $X$,
$\mathcal{C}$ is an effective curve that shrinks to a conifold along a wall of the K\"ahler cone.
From the above we learn that for any divisor class $[D]$ obeying $[D]\cdot [\mathcal{C}]<-1$,
every representative $D$ of $[D]$ necessarily has a non-trivial star-crossing singularity, and in particular is not smooth.\footnote{The condition $[D]\cdot [\mathcal{C}]<-1$ implies that $\mathcal{O}_X(D)$ is not ample, consistent with the fact that generic divisors associated with very ample line bundles are smooth.}

\subsection{The normalization of a divisor}\label{sec:defnorm}
Let $D$ be a divisor that is smooth except for a star-crossing singularity along a rational curve $\mathcal{C}$, and is birationally equivalent to a smooth divisor $\widetilde{D}$ via a flop transition of the ambient Calabi-Yau threefold.

We will now describe the divisor transition $D$ to $\widetilde{D}$ as the composition of two elementary operations. The first is the \textit{normalization} $D \leftarrow  \overline{D}$,
defined in this context\footnote{The normalization is a canonical partial desingularization process. In general, this prescription can induce and resolve singularities in various ways. However, in the present context, the normalization procedure will precisely resolve the singularities along smooth, rational intersection curves.}  as the blow-up of the relevant $\mathbb{P}^1$s on $X$. One can think of this as replacing the intersection curve $\mathcal{C}$ by $k$ disjoint copies $\mathcal{C}_{i}$, $i=1,\ldots,k$, with one glued into each of the $k$ local branches. The normalization $\overline{D}$ is thus smooth. The normalization process for a $k=3$ star-crossing singularity is shown in Fig.~\ref{fig:normalcrossing} and \ref{fig:normalization}.

Each of the curves $\mathcal{C}_{i}$ is a curve of self-intersection number $-1$ in $\overline{D}$ and can thus be blown down to a smooth point. The divisor transition $D$ to $\widetilde{D}$ is precisely the pullback to the normalization, followed by the blow-down. We note that
\begin{equation}
	h^\bullet \Bigl(\widetilde{D},\mathcal{O}_{\widetilde{D}}\Bigr)=h^\bullet\Bigl(\,\overbar{D},\mathcal{O}_{\overbar{D}}\Bigr)\, ,
\end{equation}
as both $\overline{D}$ and $\widetilde{D}$ are smooth,
and blowing down curves of self-intersection number $-1$ leaves $h^\bullet$ invariant.

Importantly, it follows that
\begin{equation}
	\smallhagal(D,\mathcal{O}_D)\equiv h^\bullet_{\pm}(\overbar{D},\mathcal{O}_{\overbar{D}})
\end{equation}
is \emph{invariant} under flop transitions. Furthermore, in the presence of star-crossing singularities with $k\geq 2$ one has
\begin{equation}\label{eq:onehas}
	\smallhagal(D,\mathcal{O}_D)\neq h^\bullet_\pm(D,\mathcal{O}_D)\, .
\end{equation}
This follows from the non-trivial transformation property \eqref{eq:chihol_trafo} of the holomorphic Euler characteristic which, for $k\geq 2$, implies $\chi(\overbar{D},\mathcal{O}_{\overbar{D}})>\chi(D,\mathcal{O}_{D})$.

\section{Zero mode counting for singular divisors from analytic continuation}\label{sec:analytic}

In the previous section we saw that flop transitions between Calabi-Yau threefolds $(X,\widetilde{X})$ can induce non-trivial geometric transitions between divisors $(D,\widetilde{D})$. In particular, we reiterate that flop transitions induce star-crossing singularities in divisors that intersect the flop curve with intersection number $\geq 2$, while they resolve star-crossing singularities for intersection numbers $\leq -2$. Using this, we will now seek a formula for fermion zero mode counting on divisors with star-crossing singularities.

\subsection{A local argument}\label{subsec:openstringargument}

We will start with a heuristic argument. Given two irreducible divisors $D$ and $D'$ that intersect each other along a curve $\mathcal{C}=D\cap D'$, one may wrap D-branes on both divisors. The zero modes living on the brane system then come from open strings starting and ending on either $D$ or $D'$. Therefore, the zero modes of $D\cup D'$ should consist of those of $D$, those of $D'$, plus, possibly, extra zero modes from strings stretching between $D$ and $D'$. The positive/negative chirality spinor bundles on $\mathcal{C}$ are equal to the line bundles $\sqrt{K}$ and $\sqrt{K}\otimes \Omega^{0,1}$, respectively, where $K$ denotes the canonical bundle on $\mathcal{C}$ and $\Omega^{0,1}$ is the bundle of $(0,1)$-forms. In the presence of worldvolume fluxes $\mathcal{F}$ on $D$ and $\mathcal{F}'$ on $D'$ the physical fermions on $\mathcal{C}$ take values in the spinor bundle of $\mathcal{C}$ twisted by the line bundle $\mathcal{L}$ with $c_1(\mathcal{L})=(\mathcal{F}-\mathcal{F}')|_{\mathcal{C}}$. Thus, the positive/negative chirality fermion zero modes localized on the intersection curve are counted by $h^0(\sqrt{K}\otimes \mathcal{L})$ and $h^1(\sqrt{K}\otimes \mathcal{L})$, respectively, with chiral index
\begin{equation}
	\chi_{\text{chiral}}=\chi(\mathcal{C},\sqrt{K}\otimes \mathcal{L}):=h^0(\sqrt{K}\otimes \mathcal{L})-h^1(\sqrt{K}\otimes \mathcal{L})=\int_{\mathcal{C}}c_1(\mathcal{L})\, .
\end{equation}
As a consequence, for trivial worldvolume fluxes the spectrum of zero modes living at the intersection curve is non-chiral. Moreover, if the intersection curve has genus $g=0$ we have $\sqrt{K}=\mathcal{O}(-1)$, which satisfies $h^i(\sqrt{K})=0$. Then, there arise \emph{no} zero modes at all from the intersection curve. We thus conclude that for a pair of divisors intersecting each other transversely along a rational curve, the physical zero modes are simply those of $D$ and those of $D'$.\footnote{Note that if the intersection curve has genus $g>0$, then generically additional fermion zero modes will be generated. For this reason we focus our analysis on rational intersection curves.} This means that the fermion zero modes are counted by those of the normalization $\overline{D\cup D'}$. It would be surprising if the localized modes from open strings stretching between the two components were sensitive to whether or not the divisor is globally the union of two irreducible components or not, so one would expect that if a divisor $D$ has \emph{local} star-crossing singularities along rational curves then the zero modes of its normalization $\overline{D}$ should still count the physical zero modes. Note that this conclusion also holds in the presence of non-trivial worldvolume flux on $D$, because $\mathcal{F}=\mathcal{F}'$ if $D$ is irreducible. Hence, we arrive at our main claim that $\smallhagal(D,\mathcal{O}_D)$ counts physical zero modes.

Thus, given $k$ smooth and irreducible divisors $D_i$ intersecting each other along rational curves,
their union $D:=\cup_iD_i$ obeys
\begin{equation}
	{\smallhagal}(D,\mathcal{O}_D)=\sum_i {\smallhagal}(D_i,\mathcal{O}_{D_i})\,.
\end{equation}
This makes sense from a physical standpoint, because we expect that each component $D_i$ carries its own independent fermion zero modes, without additional zero modes along the intersection loci, at least in the absence of a non-trivial gauge bundle.

\subsection{An analytic continuation argument}\label{subsec:holomorphy}

We can further substantiate our claim by using the fact that Euclidean D3-branes wrapped on effective divisors contribute to superpotential-like interactions in the effective theory. By superpotential-like, we mean terms in the effective action of the form $\int d^2\theta\ldots$ that cannot be written as $\int d^4\theta \ldots $, but we also allow superspace derivatives in the argument.
One should be able to analytically continue holomorphic couplings throughout K\"ahler moduli space, at least in regimes where all the instanton actions remain large.

Importantly, holomorphy forbids any contributions to the superpotential (or more generally superpotential-like interactions) from Euclidean $(p,q)$ strings wrapped on curves. As a consequence, no relevant instanton action becomes small upon undergoing a flop transition. In fact, given a facet of the K\"ahler cone corresponding to a flop, we can scale up the K\"ahler parameters along the facet to make divisor volumes (and thus the relevant instanton actions) arbitrarily large.

Thus, given a contribution to
a superpotential-like interaction from a Euclidean D3-brane wrapped on a divisor $D$, one expects that its Pfaffian $\mathcal{A}_D(z,\tau)$ should in fact be invariant under flop transitions, and in particular the zero mode counting should remain invariant.\footnote{There is a slight subtlety with this argument: given a flop transition involving the shrinking of a curve that intersects O7-planes non-trivially, the Freed-Witten anomaly \cite{Freed:1999vc} implies a half-integer B-field along that curve, which implies that one can analytically continue from one phase to the next without encountering a singularity (see e.g. \cite{Demirtas:2021nlu} for a discussion of this). If, however, the shrinking curve does not intersect O-planes it might not be possible to choose a half-integral B-field to avoid the singularity. However, even in this case, one should always be able to avoid the singularity by perturbing an orientifold-compatible non-zero mode of the B-field by an arbitrarily small but non-zero amount to move around the singularity.} Therefore, as long as there exists at least one geometric phase in which $D$ is smooth (the divisors considered in the examples in \S \ref{sec:examples} satisfy this property), the zero mode counting should be the standard one applied in that particular phase. Moreover, since $D$ is smooth in said phase, it is equal to its normalization, and therefore the zero modes are counted by $\smallhagal(D,\mathcal{O}_D)$ in that phase. But as we have seen that $\smallhagal(D,\mathcal{O}_D)$ is invariant under flops, it follows that $\smallhagal(D,\mathcal{O}_D)$ counts zero modes in \textit{any} phase.

\subsection{Superpotentials and symmetric flop transitions}

Before we discuss   concrete examples we would like to make one more point. As explained in \S \ref{sec:symm_flops}, a flop transition connecting two geometric phases that are diffeomorphic to one another is called a \textit{symmetric flop transition} \cite{Brodie:2021ain}, and we will consider Calabi-Yau compactifications admitting infinite series of symmetric flop transitions.
Although a fundamental domain of moduli space can be obtained by restricting to any one of the diffeomorphic phases, the flop transitions carry important topological information, because they induce non-trivial  monodromies  of the curve and divisor lattices upon continuing from one phase to the next. The monodromies can be thought of as linear transformations $\Lambda$ acting as
	\begin{align}	
	&H_4(X,\mathbb{Z})\rightarrow H_4(X,\mathbb{Z})\, ,\quad \vec{Q}\mapsto \Lambda\cdot \vec{Q}\, \\
    &H_2(X,\mathbb{Z})\rightarrow H_2(X,\mathbb{Z})\, ,\quad \vec{q}\mapsto (\Lambda^{-1})^T\cdot \vec{q}\,.
	\end{align}
Because the monodromy relates different effective divisor classes, it can have important implications for instanton physics. In particular, let us assume that a given divisor $D$ contributes to the nonperturbative superpotential as evaluated in some geometric phase.

Then, as in \S\ref{subsec:holomorphy}, by analytic continuation through the flop transition $X\rightarrow X'\overset{\Lambda}{\simeq} X$, we learn that a Euclidean D3-brane wrapped on the divisor class $\Lambda \cdot \vec{Q}$ must contribute to the EFT of the phase $X'$ in \textit{precisely the same way} that a Euclidean D3-brane on $\vec{Q}$ contributes in $X$: at least, these contributions must be equal in the absence of backgrounds that spontaneously break the monodromy symmetry. This observation has particularly interesting consequences if the monodromy group has infinite order. In this case, the superpotential must contain an infinite number of terms
\begin{equation}
	W(T)\supset \mathcal{A}_{D}(z,\tau)\sum_{\ell \in \mathbb{Z}}e^{-2\pi \int_{\Lambda^\ell (D)}\frac{1}{2}J\wedge J -iC_4}\, .
\end{equation}
In our examples in \S\ref{sec:examples} we will see that such sums can give rise to Jacobi forms.

\section{Modular superpotentials}\label{sec:examples}

In this section we discuss two explicit examples of Calabi-Yau orientifolds that feature non-trivial (and in fact infinitely many!) contributions to the superpotential from Euclidean D3-branes wrapping star-rigid divisors with star-crossing singularities. These terms can be summed into Jacobi theta functions, and thus have interesting modular transformation properties.

	\subsection{Example 1}\label{sec:example1}
	
	We consider the anticanonical hypersurface in a toric variety $V$ defined by a toric fan whose elements are the cones over the simplices of a fine regular star triangulation (FRST) $\mathcal{T}$ (ignoring points interior to facets) of a reflexive polytope $\Delta^\circ\subset  N_{\mathbb{R}}:=N\otimes \mathbb{R}$ that is the convex hull of a set of lattice points in $N\simeq \mathbb{Z}^4$. The points in $\Delta^\circ\cap N$ not interior to facets are the  origin, and the columns of
	\begin{equation}
	\begin{pmatrix}
	-1 & -1 & 0  & -1 & -1 & 1  & 1\\
	-1 & 1  & -1 & -1 & -1 & 1  & 1\\
	0  & -1 & 1  & 0  & 1  & -1 & 0\\
	0  & -1 & 1  & 1  & 0  & 0  & -1
	\end{pmatrix}\, .
	\end{equation}
	A GLSM charge matrix is given by
	\begin{equation}\label{eq:GLSM}
	Q=\begin{pmatrix}
		0& 0& 0& 1& 0& 0& 1\\
		0& 0& 0& 0& 1& 1& 0\\
		1& 1& 2& -1& -1& 0& 0
	\end{pmatrix}\, ,
	\end{equation}
	and one may choose $\mathcal{T}$ such that the K\"ahler cone conditions are $t^i>0$, where the $t^i$ are the Fayet-Iliopoulos parameters associated with the rows of $Q$. In this case, the Stanley-Reisner (SR) ideal is generated by $\{x_1x_2x_3,x_4x_7,x_5x_6\}$.
	
	Equation \eqref{eq:GLSM} identifies $V$ as a $\mathbb{P}^1\times \mathbb{P}^1$ fibration over the base $\mathbb{P}^2_{112}$, and in particular it admits two distinct $\mathbb{P}^1$ fibrations. Over each point in $\mathbb{P}^2_{112}$, setting $x_4=0$ defines a point in the first $\mathbb{P}^1$, while $x_5=0$ defines a point in the second $\mathbb{P}^1$. Hence, the intersection surface $D_4\cap D_5$ is equivalent to the base of the fibration. We note a symmetry exchanging the two $\mathbb{P}^1$ factors, acting on the curve basis $\{\mathcal{C}^1,\mathcal{C}^2,\mathcal{C}^3\}$ by permuting $\mathcal{C}^1\leftrightarrow \mathcal{C}^2$ and permuting the prime toric divisors as
	\begin{equation}
	D_4\leftrightarrow D_5\, ,\quad D_6\leftrightarrow D_7\, .
	\end{equation}
	We choose the divisor classes $(H_1,H_2,H_3):=([D_7],[D_6],[D_1])$ as a basis of $H_6(V,\mathbb{Z})$ such that the columns of $Q$ correspond to the charges of the prime toric divisors, and we expand the K\"ahler form $J$ as
	\begin{equation}
	J=\sum_{a=1}^3t^aH_a\, .
	\end{equation}
	In the limit $t^{1}\rightarrow 0$ the first $\mathbb{P}^1$ factor of the fibration shrinks to zero size, while in the limit $t^{2}\rightarrow 0$ the second factor shrinks. In the limit $t^3\rightarrow 0$ the base shrinks. Therefore, the K\"ahler cone $\mathcal{K}_V$ of the ambient variety $V$ is given by
	\begin{equation}
	t^a>0\, ,\quad a=1,2,3\, .
	\end{equation}
	At each of the three walls, effective divisors, surfaces, and curves shrink to zero volume.
	
	Next, we consider the anticanonical hypersurface $X \subset V$, which is a Calabi-Yau threefold. Intersecting the $H_a$ with $X$, we obtain a basis of $H_4(X,\mathbb{Z})$, and the independent triple intersection numbers $\kappa_{abc}:=H_a\cdot H_b\cdot H_c$ are
	\begin{equation}\label{eq:tripleint1}
		\mathcal{K}_{1ab}=\begin{pmatrix}
			1& 3& 1 \\
			& 3& 3 \\
			& & 1
		\end{pmatrix}\, ,\quad \mathcal{K}_{2ab}=\begin{pmatrix}
			1& 1 \\
			& 1
		\end{pmatrix}\, ,\quad \mathcal{K}_{333}=0\, ,
	\end{equation}
	where we only display the $\kappa_{abc}$ with $a\leq b\leq c$.
	
	First, one can show that none of the effective varieties in $V$ whose volumes shrink at the walls of $\mathcal{K}_V$ intersect the Calabi-Yau $X$ in shrinking subvarieties. Therefore, a priori, the K\"ahler cone $\mathcal{K}_X$ of $X$ could be bigger than that of $V$. However, using mirror symmetry \cite{Candelas:1990rm,Hosono:1993qy,Hosono:1994ax,compmirror} one can compute the Gopakumar-Vafa (GV) invariants \cite{Gopakumar:1998ii,Gopakumar:1998jq} of the basis curves of the Mori cone inherited from $V$, giving
	\begin{equation}
	n^0_{(1,0,0)}=48\, ,\quad n^0_{(0,1,0)}=48\, ,\quad n^0_{(0,0,1)}=1\, .
	\end{equation}
	Thus, all three generators are effective in $X$ and one actually has $\mathcal{K}_X=\mathcal{K}_V$.

	Using the intersection form, one sees that the curve class $[\mathcal{C}]=(0,0,1)$ can be represented by the complete intersection $D_4\cap D_5\subset X$. This identifies the curve as a weight two hypersurface in the base $\mathbb{P}_{[1,1,2]}$, i.e.~a single $\mathbb{P}^1$, consistent with its GV invariant. Its normal bundle is
	\begin{equation}
		\Bigl(\mathcal{O}_X(D_4)\oplus \mathcal{O}_X(D_5)\Bigr)\Bigl|_{\mathcal{C}}\simeq \mathcal{O}_{\mathbb{P}^1}(-1)\oplus \mathcal{O}_{\mathbb{P}^1}(-1)\, .
	\end{equation}
	Thus, $\mathcal{C}$ shrinks to a conifold as $t^3\rightarrow 0$.
	
	The other two generators of the Mori cone cannot be written as complete intersections in $X$. However, one can study them as codimension-three varieties in $V$. To see how this works for the curve class $(1,0,0)$, consider writing the Calabi-Yau defining equation as
	\begin{equation}
		f=A_0 x_4^2+A_1 x_4x_7 +A_2 x_7^2\, ,
	\end{equation}
	where the $A_{i}$ are generic sections of $\mathcal{O}_V(2H_2+(4-i)H_3)$. The $A_i$ do not depend on the projective $\mathbb{P}^1$ coordinates $[x_4:x_7]$, so the codimension-three locus defined by $A_0=A_1=A_2=0$ is a disjoint set of copies of the $\mathbb{P}^1$ fiber parameterized by $[x_4:x_7]$. As $f\equiv 0$ along this locus, this set of $\mathbb{P}^1$'s is contained in the Calabi-Yau. Moreover, the $A_i$ can be viewed as generic sections of line bundles on the base of the $\mathbb{P}^1$ fibration, and it is therefore a straightforward exercise in intersection theory to confirm that there are $48$ $\mathbb{P}^1$'s, as predicted by the GV invariant computed via mirror symmetry.  Similarly, using that $[x_5:x_6]$ are projective coordinates on the other $\mathbb{P}^1$ fiber, one shows that there are $48$ isolated $\mathbb{P}^1$'s in the class $(0,1,0)$. Near each isolated $\mathbb{P}^1$, we can define a local patch $\simeq \mathbb{C}^3\times \mathbb{P}^1$, and use $(A_0,A_1,A_2)$ as coordinates along $\mathbb{C}^3$. With this local description, one can use standard algebraic methods, implemented in {\tt{Macaulay2}} \cite{M2}, to show that the normal bundle of each isolated $\mathbb{P}^1$ is $\mathcal{O}(-1)\oplus \mathcal{O}(-1)$. Therefore, the local geometry around each of the vanishing curves is again isomorphic to the resolved conifold.
	
	One arrives at the same conclusion via the following alternative route. We consider the wall of $\mathcal{K}_X$ where the curve with charge $(0,0,1)$ shrinks. Using the intersection form \eqref{eq:tripleint1} we find
	\begin{equation}
		\int_{H_a}\frac{1}{2}J\wedge J \quad \overset{t^3\rightarrow 0}{\longrightarrow} \quad \frac{1}{2}A\cdot \begin{pmatrix}
		(t^1)^2\\
		t^1t^2\\
		(t^2)^2
	\end{pmatrix}\, ,\quad
A:=\begin{pmatrix}
	1& 6 & 3\\
	3& 6 & 1\\
	1& 6 & 1
\end{pmatrix}\,.
	\end{equation}
	As $\det A=24\neq 0$, there cannot exist any effective divisor class $q^aH_a$ with the property that its calibrated volume vanishes identically along the locus where the curve with charge $(0,0,1)$ shrinks. Hence, recalling the discussion of \S\ref{sec:flop}, we conclude that continuing past this wall of the K\"ahler cone is a flop transition.

	Next, starting in the phase $X$, we consider passing through the wall where the curve with charge $(1,0,0)$ shrinks. In this case, one sees that the unique divisor class $\mathcal{D}$ that satisfies $\int_{\mathcal{D}}\frac{1}{2}J\wedge J|_{t^1=0}=0$, and $\int_{\mathcal{D}}\frac{1}{2}J\wedge J|_{t^1>0}>0$, for all $t^{2,3}>0$, is
	\begin{equation}\label{eq:defdn}
		\mathcal{D}:=(-1,1,2)\equiv -H_1+H_2+2H_3=-D_7+D_6+2D_1\,.
	\end{equation}
	Thus, if $\mathcal{D}$ were effective, it would be the unique divisor class that shrinks in the limit $t^1\rightarrow 0$,
	\begin{equation}
		\text{Vol}(\mathcal{D})=\int_{\mathcal{D}} \frac{1}{2}J\wedge J=2 t^1 (t^1 + 3 t^2 + 2 t^3)\overset{t^1\rightarrow 0}{\longrightarrow} 0\, .
	\end{equation}
However, using the methods of Appendix B of \cite{Demirtas:2019lfi}, one can show that
$\mathcal{D}$ is \textit{not} effective. Thus, we conclude that continuing through this wall of the K\"ahler cone also constitutes a flop transition.

We have now established in two separate ways that the walls of $\mathcal{K}_X$ are flop walls, and we can proceed to discuss the neighboring phases.
On the other side of the wall where $t^3 \rightarrow 0$, we encounter another geometric phase $X'$.  Its Mori cone generators have charges $(0,1,1)$, $(1,0,1)$ and $(0,0,-1)$, and their GV invariants are
\begin{equation}
	n^0_{(1,1,0)}=56\, ,\quad n^0_{(0,1,0)}=56\, ,\quad n^0_{(0,0,-1)}=1\, .
\end{equation}
At the two walls of $\mathcal{K}_{X'}$ that correspond to the shrinking of the first two generators, the prime toric divisors $D_{4}$ and $D_5$, respectively, shrink to points, while the overall volume of $X$ stays finite. Thus, these two walls are facets of the extended K\"ahler cone. The third generator can be flopped to return to the phase $X$.

Beyond the wall where $t^1 \rightarrow 0$, we again encounter a geometric phase $X''$.  Its Mori cone generators have charges $(2,1,0)$, $(-1,0,0)$ and $(4,0,1)$ and GV invariants
	\begin{equation}
	n^0_{(2,1,0)}=48\, ,\quad n^0_{(-1,0,0)}=48\, ,\quad n^0_{(4,0,1)}=1\, .
	\end{equation}
	Note that these are the same as the GV invariants of the generators of $X$. Indeed, one may compute the transformed triple intersection numbers and second Chern class, using \eqref{eq:tripleintC2_trafo},
	and express them in a basis of divisors dual to the generators of the new Mori cone.  One finds that indeed the new triple intersection numbers and second Chern class are exactly the same as those of $X$ computed in our original basis.
	
\begin{figure}
	\centering
	\includegraphics[width=0.4\textwidth]{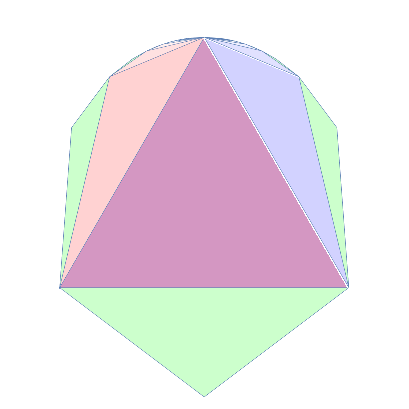}
	\caption{A projection of the three-dimensional extended K\"ahler cone, $\mathcal{K}$, for Example $1$. The center region is $\mathcal{K}_X$, the K\"ahler cone of the original toric phase, $X$. The green regions represent flops that map $X$ to a birationally equivalent, but non-isomorphic phase, while the pastel rose and lavender regions represent flops to isomorphic phases. At the outer boundaries, $D_4$, $D_5$, and their infinite set of images under $ \Lambda^{\ell}$ shrink to a point.}
	\label{fig:ply45extk}
	\end{figure}

	Hence, $X\simeq X''$, i.e.~we encounter a symmetric flop. The linear transformation that maps the generators of the Mori cone of $X$ to those of $X''$ is
	\begin{equation}
	(\Lambda^{-1})^T=\begin{pmatrix}
	2 & -1 & 4\\
	1 & 0 & 0\\
	0 & 0 & 1
	\end{pmatrix}\, .
	\end{equation}
	As $\Lambda^\ell\neq \id$ for all $\ell\neq 0$, we have uncovered an infinite series of symmetric flops where the Mori cone generators of the $\ell$-th phase are obtained by multiplying the generators of the Mori cone of $X$ by the matrix $\left(\Lambda^{-\ell}\right)^T$. Similarly, we obtain an infinite series of effective divisors by applying $\Lambda^\ell$ to the columns of $Q$. The result is
	\begin{equation}
	\Lambda^\ell\cdot Q=
	\begin{pmatrix}
	0& 0& 0& 1 - \ell& -\ell& -\ell& 1 - \ell&\\
	0& 0& 0& \ell& 1 + \ell& 1 + \ell& \ell\\
	1& 1& 2& -1 - 2 \ell + 2 \ell^2& -1 + 2 \ell + 2 \ell^2& +2 \ell + 2 \ell^2& -2 \ell + 2 \ell^2
	\end{pmatrix}\, .
	\end{equation}
	The divisors $\Lambda^\ell\cdot Q$ take values in a cone whose extremal generators are the images of $D_4$ under the monodromy group generated by $\Lambda$, i.e.
	\begin{equation}\label{eqn:image divisor example 1}
	D^{(\ell)}:=(1-\ell,\ell,-1-2\ell+2\ell^2)\, ,
	\end{equation}
	with $D^{(0)}\equiv D_4$ and $D^{(1)}\equiv D_5$. Starting from each symmetric image of $X$ one can also flop to a phase isomorphic to $X'$, and the divisors that shrink at the outer walls of these new phases are precisely the $D^{(\ell)}$ with appropriate $\ell$. Therefore, we find that the effective cone of $X$ (and all flopped phases) is
	\begin{equation}
	\mathcal{E}(X)=\text{span}_{\mathbb{R}_+}\bigl\{D^{(\ell)}\bigr\}_{\ell\in \mathbb{Z}}\, .
	\end{equation}
	In particular, the divisors $D^{(\ell)}$ are all irreducible. A projection of the three-dimensional extended K\"ahler cone, $\mathcal{K}$, is shown in Fig.~\ref{fig:ply45extk}. All divisors in this series except $D_4$ and $D_5$ are autochthonous, i.e.~they are not inherited from intersecting effective divisors of the toric ambient variety $V$ with the Calabi-Yau hypersurface. Furthermore, there are four autochthonous divisors, namely $D^{(-1)}$, $D^{(2)}$, $\Lambda^{-1}D_7$, and $\Lambda^{2}D_7$, that are easily seen to be complete intersection surfaces in $V$ along which the defining polynomial of $X$ vanishes identically. To find these divisors we rewrite the defining polynomial $f$ as
	\begin{equation}
		f=x_4g_1^{(1)}+x_7^2g_2^{(1)} =x_4^2g_1^{(2)}+x_7g_2^{(2)}=x_5^2g_1^{(3)}+x_6g_2^{(3)}=x_5 g_1^{(4)}+x_6^2 g_2^{(4)}\, ,
	\end{equation}
	where the $g_{\alpha}^{(i)}$ are generic sections of the appropriate line bundles. The four autochthonous divisors $\{D^{(2)},\Lambda^2D_7,D^{(-1)},\Lambda^{-1}D_7\}$ are the (generically smooth) codimension-two surfaces $g_1^{(i)}=g_2^{(i)}=0$, $i=1,\ldots,4$. For $\ell\notin \{-1,0,1,2\}$, with some effort one can confirm explicitly (at least for small enough $\ell$) using line bundle cohomology that the divisors are indeed effective. As $\ell$ grows, the computation using line bundle cohomology quickly becomes computationally infeasible, but happily the action of the monodromy group gives the exact effective cone directly.

	Now, $D_4$ and $D_5$ are smooth and rigid divisors in $X$, with Hodge vectors $h^\bullet(D,\mathcal{O}_D)=(1,0,0)$. As all the $D^{(\ell)}$ are related to $D_4$ via an action of the monodromy group we find
	\begin{equation}
		{\small \smallhagalbare (D^{(\ell)},\mathcal{O}_{D^{(\ell)}})=(1,0,0)\quad \forall \ell\, ,}
	\end{equation}
	i.e.~all divisors $D^{(\ell)}$ are star-rigid. In contrast,  $h^\bullet (D^{(\ell)},\mathcal{O}_{D^{(\ell)}})$ satisfies
	\begin{equation}\label{eq:hbulletexample1}
		h^\bullet (D^{(\ell)},\mathcal{O}_{D^{(\ell)}})=\bigl(1,2(\ell+1) \ell (\ell-1) (\ell-2),0\bigr)\, .
	\end{equation}
	
	Up to this point, we have been studying the topological data of an $\mathcal{N}=2$ compactification on $X$, but we now turn to constructing an orientifold of $X$. We choose an involution
	\begin{equation}
	\mathcal{I}_B:\, x_1\mapsto -x_1\, ,
	\end{equation}
	acting only on the $\mathbb{P}^2_{[112]}$ base.
	In the ambient variety, the fixed locus of $\mathcal{I}_B$ is
	\begin{align}
	\mathcal{F}_B=&\{x_1=0\}\cup \{x_2=x_4=x_5=0\}\cup \{x_2=x_4=x_6=0\} \nonumber  \\
	&\cup\{x_2=x_5=x_7=0\} \cup \{x_2=x_6=x_7=0\}\, .
	\end{align}
	The first component descends to a K3 divisor $D_1$ in $X$ (i.e.~$\chi(D_1)=24$) hosting an O7-plane, and the triple intersection curves intersect $X$ in $1+2+2+3=8$ isolated points hosting O3-planes. We can cancel the D3-brane and D7-brane tapole by putting 4 D7-branes on top of the O7-plane, giving rise to gauge algebra $\mathfrak{so}(8)$, and introduce $8$ mobile D3-branes.
	
	The divisor $D_4$ is rigid, smooth and orientifold invariant in the toric phase, and $D_4$ intersects the O7-plane divisor transversely. Therefore,
	from the condition \eqref{eq:Witten96} one concludes that Euclidean D3-branes wrapped on $D_4$ should generate a superpotential term.\footnote{One slight complication remains: the intersection curve has genus $g=1$, so one expects charged zero modes from strings stretched between the Euclidean D3-brane and the seven-brane stack.  These zero modes could make the superpotential vanish if the seven-brane gauge group remained un-Higgsed. However, such zero modes should in general be lifted when the seven-brane gauge group is fully Higgsed, e.g.~by placing the mobile D3-branes onto the seven-brane stack and going to a generic point on the Higgs branch.  We will therefore assume that a superpotential is generated by $D_4$.  In any case, this subtlety will not arise in the example in \S\ref{sec:example2}.} By holomorphy, it follows that \emph{all} the $D^{(\ell)}$ must contribute to the superpotential.

We saw above that the $D^{(\ell)}$ with $-1\leq \ell\leq 2$ are smooth, so the standard rigidity condition of \eqref{eq:Witten96} applies, and is indeed fulfilled by these divisors;
so it is not surprising that $D^{(-1)},\ldots,D^{(2)}$ support superpotential contributions.
However, the $D^{(\ell)}$ with $\ell>2$ or $\ell<-1$ have non-trivial star-crossing singularities, so \eqref{eq:Witten96} is inapplicable, and moreover, in view of \eqref{eq:onehas}, is not fulfilled.
However, all the $D^{(\ell)}$ are star-rigid, i.e.~they fulfill \eqref{eq:hagalboxtimes}.
		
Now we will explain why compactifying type IIB string theory on this geometry gives rise to a modular superpotential. We have learned that the effective cone $\mathcal{E}(X)$ is generated by the divisor classes $\left\{D^{(\ell)}\right\}_{\ell\in \mathbb{Z}}$ of \eqref{eqn:image divisor example 1}, all of which have ${\smallhagalplus}(D^{(\ell)})=(1,0,0)$, and $\smallhagalminus(D^{(\ell)})=0$. By virtue of our preceding discussion, all the $D^{(\ell)}$ contribute to the superpotential.  Moreover, there are no other star-rigid divisors, so we have identified all contributing divisors.

	It remains to be shown that there exists a D-brane configuration that leaves the monodromy symmetry unbroken: as pointed out in \cite{Braun:2018fdp}, such breaking effects can be an issue in the related construction \cite{Donagi:1996yf}.
	
	The monodromy symmetry $\Lambda$ leaves the O7-plane class invariant, and the most straightforward interpretation of this is that the orientifold involution leaves the monodromy group unbroken. Strictly speaking, by continuing from one phase to the next, the orientifold involution might transform to a different one, that also has an O7-plane on a divisor in the same class $[D_1]$. There is no obvious candidate for such an alternative orientifold, so we will assume that indeed the orientifold involution gets mapped to itself. Thus, as long as all D7-branes are placed on top of the O7-plane, the presence of seven-branes does not break the monodromy symmetry. Finally, the D3-branes can be placed on top of one of the three O3-planes in the triple intersection locus $D_2\cap D_6\cap D_7$, which, likewise, are mapped to themselves by the monodromy symmetry, and are not contained in the rigid divisors hosting Euclidean D3-branes.\footnote{The classes $[D_6]$ and $[D_7]$ are not mapped to themselves under the monodromy group, but with some effort one can show that, for an appropriate choice of sections of the images of $[D_6]$ and $[D_7]$ under the action of $\Gamma$, the three points $D_2\cap D_6\cap D_7$ are actually mapped to themselves.} Thus, in our orientifold vacuum, provided the D3-brane and D7-brane position moduli take the special values stated above, all the $D^{(\ell)}$ contribute with equal Pfaffian prefactors.

 Finally, we arrive at a nonperturbative superpotential given by
	\begin{align}\label{eq:modular_superpotential1}
	W(T,z,\tau)=&~\mathcal{A}(z,\tau)\sum_{\ell\in \mathbb{Z}}e^{-2\pi\left[(1-\ell)T_1+\ell T_2 +(2\ell^2-2\ell-1)T_3\right]}+\text{multi-wrapping} \nonumber\\
	=&~\mathcal{A}(z,\tau)e^{-2\pi\frac{T_1+T_2-3T_3}{2}}\vartheta_{10}\left(Z_-;\,\tau_3\right)+\text{multi-wrapping}\, ,
	\end{align}
	where $Z_-:=i(T_2-T_1)$, $\tau_3:=4iT_3$, and $\vartheta_{10}( z;\,\tau)=\sum_{\ell\in \mathbb{Z}}e^{\pi i (\ell-\frac{1}{2})^2 \tau +2\pi i (\ell-\frac{1}{2}) z}$ is a Jacobi theta function.  The one-loop Pfaffian $\mathcal{A}(z,\tau)$ remains as an overall unknown function of the complex structure moduli and the dilaton.

Some of the symmetries of the superpotential are generated precisely by the axion shift symmetries $T_j\mapsto T_j+ia_j$, $a_j\in \mathbb{Z}$, and by the monodromy symmetry group generated by $\Lambda$. In addition, the superpotential enjoys \textit{modular} properties\footnote{For related investigations of superpotentials with modular properties, see \cite{Donagi:1996yf,Curio:1997rn,Grimm:2007xm,Kerstan:2012cy,Anderson:2015yzz,Braun:2018fdp}.}
with respect to modular transformations of $\tau_3$, arising from the Jacobi identity
	\begin{equation}\label{eq:modtrans}
	(z,\tau)\rightarrow \left(\frac{z}{\tau},-\frac{1}{\tau}\right)\, :\quad \vartheta_{10}(z;\,\tau)\rightarrow (-i \tau_3)^{\frac{1}{2}}e^{\pi i \frac{z^2}{\tau}}\vartheta_{01}(z;\,\tau)\, .
	\end{equation}
The modular transformation symmetry \eqref{eq:modtrans} of the superpotential
suggests the existence of a rather exotic
strong/weak duality symmetry of the effective field theory, involving the inversion of divisor volumes.
	
	The contributions from Euclidean D3-branes wrapping irreducible star-rigid divisors $D^{(\ell)}$ might be only the lowest terms in a multi-covering expansion\footnote{However, see
the recent work \cite{Alexandrov:2022mmy}, which argues that multi-covering contributions are absent in this setting.} from Euclidean D3-branes wrapping $D^{(\ell)}$ arbitrary many times, as is familiar from the worldsheet expansion of the prepotential in type IIA string theory on a Calabi-Yau threefold. Even if so, we may trust the leading term proportional to the Jacobi theta function at least when all the instanton actions are large. This means that our leading term \eqref{eq:modular_superpotential1} approximates the full superpotential well as long as
	\begin{equation}
		\text{Re}(T_{1,2})>\text{Re}(T_3)> 0\, .
	\end{equation}
	In particular, for all values of $Z_-$ and $\tau_3$ we can neglect possible multi-wrapping effects provided $\text{Re}(T_1+T_2)$ is suitably large.
	
	It would be very interesting to understand the vacuum structure generated by the superpotential \eqref{eq:modular_superpotential1}. However, in general this would require knowledge of the K\"ahler potential, which is difficult to obtain in the desired regime where many (or even infinitely many) instantons become relevant. Only supersymmetric Minkowski vacua are an exception, as these arise at points in moduli space where $dW=W=0$, which is an overconstrained but holomorphic system of equations. But in our case such vacua do not exist, as Jacobi theta functions only have simple roots.
	
	\subsection{Example 2}\label{sec:example2}
	
	We now turn to our second example. Again, we construct a Calabi-Yau variety $X$ as the generic anticanonical hypersurface in a toric variety $V$. As before, $V$ is obtained via an FRST $\mathcal{T}$ of a reflexive polytope $\Delta^\circ\subset N_{\mathbb{R}}$ whose points not in the interior of a facet are the origin, and the columns of
	\begin{equation}
		\begin{pmatrix}
			1 & 1 & -1 & -1 & -1 & -1 & -1 & -1 \\
			-1 & 0 &0 &0 & 1 &1  &0 &1\\
			0 &-1 &0 &0 &1 &1 &1 & 0\\
			-1 &0&0&1&0&1&0&1
		\end{pmatrix}\, ,
	\end{equation}
	and thus $h^{1,1}(V)=h^{1,1}(X)=4$. A GLSM charge matrix is given by
	\begin{align}\label{eq:glsm2}
		Q=\begin{pmatrix}
			1 & 0 & 0 &0 &   0 &0 & 0 & 1 \\
			0 & 1 & 0 &0 &   0 &0 & 1 & 0 \\
			0 & 0 & 1 &0 &   0 &1 & -1 & -1 \\
			0 & 0 & 0 &1 &   1 &0 & -1 & -1 \\
		\end{pmatrix}\, ,
	\end{align}
	which identifies the toric fourfold as a $\mathbb{P}^1\times \mathbb{P}^1$ fibration over a $\mathbb{P}^1\times \mathbb{P}^1$ base.\footnote{The anticanonical hypersurface $X\subset V$ is therefore torus fibered, and even K3-fibered, in at least two inequivalent ways.} Each column of $Q$ is associated with a homogeneous coordinate $x_i$, $i=1,\ldots,8$, and we again define $\widehat{D}_i:=\{x_i=0\}\subset V$, and $D_i:=X\cap \widehat{D}_i$. The SR ideal is generated by the monomials
	\begin{equation}
		\{x_1x_8,x_2x_7,x_3x_6,x_7x_8,x_1x_4x_5,x_2x_4x_5\}\, .
	\end{equation}
	We will work in a basis of $H^2(V,\mathbb{Z})\simeq H^2(X,\mathbb{Z})$ given by the divisor classes $H_{1,2,3,4}:=[D_{1,2,3,4}]$, adapted to the GLSM \eqref{eq:glsm2}, and a dual basis of curves $\mathcal{C}^a$, i.e.~$\mathcal{C}^a\cdot H_b={\delta^a}_b$. One can choose $\mathcal{T}$ such that the generators of the Mori cone of the toric fourfold, expressed in our basis of curves, are the columns of
	\begin{equation}\label{eq:Moriexample2}
		\begin{pmatrix}
			0 & 1 & 0 & 0 & 1 & 1\\
			1 & 0 & 1 & 1 & 0 & 0\\
			0 & 0 & 0 & 1 & 0 & 1\\
			0 & 0 & 1 & 0 & 1 & 0
		\end{pmatrix}\, ,
	\end{equation}
	and the independent triple intersection numbers are
	\begin{equation}
		\kappa_{1ab}=\begin{pmatrix}
			4 & 8 & 2 & 2\\
			  & 8 & 4 & 4\\
			  &   & 0 & 2\\
			  &   &   & 0
		\end{pmatrix}\, ,\quad
		\kappa_{2ab}=\begin{pmatrix}
			4 & 2 & 2 \\
			  & 0 & 2 \\
			  &   & 0
		\end{pmatrix}\, ,\quad
		\kappa_{3ab}=\begin{pmatrix}
			0 & 0 \\
			  & 0
		\end{pmatrix}\, ,\quad
		\kappa_{444}=0\, .
	\end{equation}
	As in our first example in \S\ref{sec:example1}, one can compute the GV invariants of the Calabi-Yau threefold $X$ to conclude that $\mathcal{M}_X\simeq \mathcal{M}_V$, so the Mori cone of $X$ is generated by the six curves $\mathcal{C}_{1},\ldots,\mathcal{C}_6$ whose coordinates are the columns of \eqref{eq:Moriexample2}. The prime toric divisor $D_8$ shrinks to a genus $5$ Riemann surface along the facets of $\mathcal{K}_X$ where either $\mathcal{C}_3$ or $\mathcal{C}_4$ shrinks.\footnote{The fact that $D_8$ shrinks to a curve can be seen from the fact that $\text{Vol}(D_8)$ vanishes linearly in $\text{Vol}(\mathcal{C}_{3,4})$. The genus of the curve is then determined by a GV computation: $g=\frac{1}{2}n_{\mathcal{C}_3}^0+1=5$.} Likewise the divisor $D_7$ shrinks along the walls of $\mathcal{K}_X$ where either $\mathcal{C}_5$ or $\mathcal{C}_6$ shrinks. Thus, these walls are facets of the extended K\"ahler cone, and the divisors $D_{7,8}$ are generators of the effective cone, associated with non-abelian enhancement to $\mathfrak{su}(2)$ with $5$ adjoint hypermultiplets.
	
	Next, we consider the asymptotic one-parameter scaling limit
	\begin{equation}\label{eq:scaling_limit}
		(t^1,t^2,t^3,t^4)\rightarrow (\lambda t^1,\lambda t^2,\lambda^{-1}t^3,t^4)\, ,\quad \lambda\rightarrow 0\, .
	\end{equation}
	The point $\lambda=0$ lies at infinite distance in moduli space, and we have
	\begin{equation}
		\text{Vol}(D_{1,2,7,8})\sim \frac{1}{\lambda}\rightarrow \infty\, ,\quad \text{Vol}(D_{3,6})\sim \lambda\rightarrow 0\, ,\quad \text{Vol}(D_{4,5})=\text{finite}\, ,
	\end{equation}
	and the overall volume $\text{Vol}(X)$ remains finite as well. We have $[D_3]=[D_6]$, so we learn that the vanishing class $[D_3]$ is also a generator of the effective cone. Similarly, by considering the scaling limit \eqref{eq:scaling_limit} with $t^3\leftrightarrow t^4$, one learns that $[D_4]$ is a generator of the effective cone. We note that $X$ can be thought of as a K3 fibration over $\mathbb{P}^1$, by interpreting either $[x_3:x_6]$ or $[x_4:x_5]$ as coordinates on the base. The K3 fibers are themselves torus fibered, and in the above limits, the generic torus fiber of the generic K3 fiber collapses, while the base of the K3 fibration blows up in such a way that the overall volume remains finite.\footnote{See \cite{Lee:2019wij} for an exploration of scaling limits of this kind.}
	
	Next, using similar reasoning as in \S\ref{sec:example1}, one learns that the remaining two walls of the K\"ahler cone, where either $\mathcal{C}^1$ or $\mathcal{C}^2$ shrinks, give rise to flop transitions.\footnote{The fact that sending $\mathcal{C}^{1,2} \rightarrow - \mathcal{C}^{1,2}$ gives rise to a flop, rather than an $\mathfrak{su}(2)$ enhancement, depends on whether any effective divisors shrink along the vanishing loci of these curves. No such effective divisor exists.} The relevant GV invariants are
	\begin{align}
		n_{\mathcal{C}_1} = 48  \ \ \ \text{and} \ \ \ n_{\mathcal{C}_2} = 48.
	\end{align}
	 Flopping either of these curves gives rise to a Calabi-Yau that is isomorphic to the original geometry. One verifies that the transformation \eqref{eq:tripleintC2_trafo} is undone by a basis change parameterized by
	\begin{align}
		(\Lambda^{-1})^T =
		\begin{pmatrix}
			0 & 1 & 0 &0  \\
			-1& 2 & 2 &2 \\
			0 & 0 & 1 &0  \\
			0 & 0 & 0 &1  \\
		\end{pmatrix}
	\end{align}
	Again, we find that $\Lambda^\ell \neq \mathbb{I}$ for any $\ell\neq 0$, and we arrive at an infinite series of flops.
	
	The Calabi-Yau defined by \eqref{eq:glsm2} has two prime toric divisors that
	are rigid in the sense of \eqref{eq:Witten96}:
	\begin{align}
		D_7 = \begin{pmatrix} 0 \\1 \\ -1 \\-1 \end{pmatrix} \ \ \text{and} \ \ D_8 = \begin{pmatrix} 1 \\0 \\ -1 \\-1 \end{pmatrix}.
	\end{align}
	As in our first example, acting on $D_7$ with $\Lambda$ gives $D_8$, and acting with further powers of $\Lambda$ gives rise to an infinite series of star-rigid autochthonous divisors.
	We find
	\begin{align}
		\Lambda^{\ell} D_7 = \begin{pmatrix} \ell \\ 1-\ell \\ \ell(\ell-1) -1 \\ \ell(\ell-1) -1 \end{pmatrix}.
	\end{align}
	In order to specify a four-dimensional $\mathcal{N}=1$ supersymmetric solution, we define the orientifold involution
	\begin{align}
		\mathcal{I}_B : x_3 \rightarrow -x_3\, ,
	\end{align}
	and find again that the orientifold is mapped to itself by the symmetric flop transition. The fixed locus is
	\begin{align}
		\mathcal{F}_B &= \{x_3 =0 \} \cup \{x_1 = x_2 = x_6 =0 \}  \nonumber \\ & \cup \{x_1 = x_6 = x_7 =0 \} \cup \{x_2 = x_6 = x_8 =0 \}.
	\end{align}
	The first component is a K3 fiber, i.e.~$\chi(D_3) = 24$, and the triple intersection curves intersect the Calabi-Yau in $4+2+2=8$ isolated points. Therefore, the divisor $D_3$ hosts an O7-plane and there are eight isolated O3-planes. The D3-brane tadpole is $Q_{D3}=8$, which can be canceled by including eight mobile D3-branes. Placing four D7-branes on top of the O7-plane, and placing the mobile D3-branes onto the locus $\{x_1 = x_2 = x_6 =0 \}$, then ensures that the monodromy group remains unbroken.
	
	As in our first example in \S\ref{sec:example1}, we now argue that an infinite series of Euclidean D3-branes wrapping $\Lambda^{\ell} D_7$ contribute to the superpotential. It suffices to show that the smooth divisor $D_7$, which acts as a seed for the infinite series, has the right number of zero modes. We have $h^\bullet_{+}(D_7,\mathcal{O}_{D_7})=(1,0,0)$ and $h^\bullet_{-}(D_7,\mathcal{O}_{D_7})=0$, so the criterion \eqref{eq:Witten96Threefold} is fulfilled. The divisor $D_7$ intersects the divisor $D_3$, which hosts the O7-plane, transversely along a $\mathbb{P}^1$.\footnote{The fact that the intersection curve is a $\mathbb{P}^1$ can be seen from the fact that the genus of the dual two-face of the polytope is zero.
	Alternatively, from the GLSM, it is easy to see that the ambient variety is a $\Bbb{P}^1\times\Bbb{P}^1$ fibration over $\Bbb{P}^1\times \Bbb{P}^1$, and that the Calabi-Yau is an elliptic fibration over $\Bbb{P}^1\times\Bbb{P}^1$. Because of this topology, $D_3$ in the Calabi-Yau is an elliptic fibration over $\Bbb{P}^1$ (i.e.~an elliptic K3) and $D_7$ is $\Bbb{P}^1\times\Bbb{P}^1$ ($\chi(D_7)=4$).  One thus sees that $D_3\cap D_7$ is a $\Bbb{P}^1$.} Thus, in contrast to our first example in \S\ref{sec:example1}, there are no additional charged zero-modes from $3-7$ strings, and we conclude that Euclidean D3-branes on $D_7$ contribute to the superpotential, even when the seven-brane gauge algebra $\mathfrak{so}(8)$ remains unhiggsed. As a consequence, all of the $\Lambda^\ell D_7$ contribute to $W$.
	
By reasoning that is completely analogous to that of \S\ref{sec:example1}, the superpotential is now  determined, up to a single overall one-loop Pfaffian:
	\begin{align}
		W = \mathcal{A}_D(z,\tau) \sum_{\ell \in \mathbb{Z}}e^{-2\pi \left(\ell T_1 + (1-\ell) T_2 + (\ell^2-\ell-1)(T_3+T_4)\right)} +\text{multi-wrapping}.
	\end{align}
	This sum can again be written in terms of a Jacobi-form:
	\begin{equation}
		W = \mathcal{A}_D(z,\tau) e^{-2\pi \left(\frac{T_1+T_2}{2}-\frac{5}{4}(T_3+T_4)\right)} \vartheta_{10}\left(Z_-,\tau_{34}\right) +\text{multi-wrapping}\, ,
	\end{equation}
	with $Z_-:=i(T_1-T_2)$, and $\tau_{34}:=2i(T_3+T_4)$.

	\pagebreak

	\section{Conclusions}\label{sec:conclusions}
Understanding nonperturbative contributions to the effective field theories that result from compactifications of string theory is a crucial step in characterizing the low-energy quantum gravity landscape. In type IIB flux compactifications on orientifolds of Calabi-Yau threefolds,
Euclidean D3-branes wrapping holomorphic four-cycles can in principle contribute to the superpotential, but determining whether a particular four-cycle supports a nonvanishing contribution requires a counting of fermion zero modes.
For the case of Euclidean D3-branes wrapping a smooth effective divisor $D$, Witten presented a counting of fermion zero modes in terms of the cohomology of the structure sheaf $\mathcal{O}_D$, and identified a sufficient condition for a superpotential contribution  \cite{Witten:1996bn}.
However, effective divisors in Calabi-Yau threefolds are very often singular, and a prescription for counting zero modes in this general case was lacking.

In this work, we obtained the sufficient condition \eqref{eq:hagalboxtimes} for superpotential contributions from Euclidean D3-branes wrapping divisors with a very common type of singularity.
Specifically, we studied divisors with singularities along rational curves, which we termed \textit{star-crossing singularities}, that are induced by passing through flop transitions: see Figure~\ref{fig:normalcrossing}.
Such singularities can be ``unwound" by performing a series of flops to a geometry that is birationally equivalent to the original Calabi-Yau, and in which the initially singular divisor becomes smooth. We argued that throughout such flops, the zero modes of Euclidean D3-branes wrapping effective divisors are tracked by holomorphic objects such as the superpotential, and therefore the number of physical zero modes remains invariant. As a consequence, a divisor $D$ with star-crossing singularities contributes to the superpotential in a particular Calabi-Yau if there exists a  birationally-equivalent geometry in which the image of $D$ is smooth and fulfills the condition of \cite{Witten:1996bn}.   We showed that the corresponding zero modes can be counted in terms of the cohomology of the structure sheaf of the \emph{normalization}
of $D$, leading to the condition  \eqref{eq:hagalboxtimes}.

We applied this prescription to two explicit examples of Calabi-Yau geometries. Within these compactifications, we showed that an infinite number of Euclidean D3-branes wrapping irreducible effective divisors contributed to the superpotential. In both examples, only four of the divisors in this infinite series were smooth and rigid in the sense of \eqref{eq:Witten96}. The rest of the divisors contained star-crossing singularities and were \textit{star-rigid}, in the sense of \eqref{eq:hagalboxtimes}. Adding up these contributions, we found that the superpotential took the form of a Jacobi theta function, whose modular transformation property suggests the existence of a new symmetry involving the inversion of divisor volumes.

\section*{Acknowledgements}

We thank Callum Brodie, Andrei Constantin, Andre Lukas, and Fabian Ruehle for sharing a preliminary draft of their related work \cite{Brodie:2021toe}.
We are grateful to Mehmet Demirtas, Jim Halverson, Ben Heidenreich, Richard Nally, Andres Rios-Tascon, and Tom Rudelius for discussions of similar topics, and to Cody Long for collaboration at an early stage of this work.  We thank Andres Rios-Tascon for providing Figure \ref{491_fan}.  
The work of N.G., L.M., and J.M.~was supported in part by NSF grant PHY-1719877. The work of M.K.~was supported by a Pappalardo Fellowship. The work of M.S. was supported by NSF DMS 20-01367.

\bibliography{biblio}
\bibliographystyle{JHEP}
\end{document}